# Possible applications of $Mo_2C$ in the orthorhombic and hexagonal phases explored via ab-initio investigations of elastic, bonding, optoelectronic and thermophysical properties

M.I. Naher, S.H. Naqib*
Department of Physics, University of Rajshahi, Rajshahi 6205, Bangladesh
*Corresponding author; Email: salehnaqib@yahoo.com

**Abstract**
Binary carbides demonstrate attractive set of physical properties that are suitable for numerous and diverse applications. In the present study, we have explored the structural properties, electronic structures, elastic constants, acoustic behaviors, phonon dispersions, optical properties, and various thermophysical properties of binary ortho- and hexa-$Mo_2C$ (O-MC and H-MC, respectively) compounds in details via first-principles calculations using the density functional theory (DFT). The calculated ground state lattice parameters in both the symmetries are in excellent agreement with available experimental results. The calculated electronic band structure, density of states, and optical properties of $Mo_2C$ in both structures reveal metallic features. The orthorhombic crystal shows higher level mechanical and thermal anisotropy compared to that in the hexagonal phase. The elastic constants and phonon dispersion calculations show that, in both structures, $Mo_2C$ is mechanically and dynamically stable. A comprehensive mechanical and thermophysical study shows that both phases possess high structural stability, reasonably good machinability, ductile nature, high hardness, low compressibility, high Debye temperature and high melting temperature. Moreover, the electronic energy density of states, electron density distribution, elastic properties, and Mulliken bond population analyses indicate that the structures under consideration consist of mixed bonding characteristics with ionic and covalent contributions. Investigation of the optical properties reveals that the reflectivity spectra are anisotropic with respect to the polarization directions of the electric field in the visible to mid-ultraviolet regions. High reflectivity over wide spectral range makes the compound suitable as reflecting coating. Both the structures are efficient absorber of ultraviolet radiation. The refractive indices are quite high in the infrared to visible range. Both structures show directional (plane) optical anisotropy. Though, hexa-$Mo_2C$ exhibits stronger optical anisotropy than the ortho-$Mo_2C$.

## 1. Introduction

Binary carbides have attracted great deal of interest in materials science due to their unique physical and chemical properties such as excellent structural stability, extreme hardness, high melting temperature, outstanding electrical and thermal conductivities, and excellent corrosion and wear resistances [1-3]. These desirable properties make them suitable for wide usages in cutting tools and wear resistance parts of machineries which need to work under high temperatures and pressures. Molybdenum carbide ($Mo_2C$), one of the most important transition metal carbides discovered till date, has demonstrated promise in thermoelectricity [4,5], electrochemical catalysis [6-9], and energy storage devices [10-12]. They also have huge potential as diffusion barriers and use in electrical connections in microelectronics [13,14]. $Mo_2C$ exhibits several outstanding features, namely, excellent catalytic activity, superconductivity and high capacity as an anode material to be employed in Li or Na-ion batteries, with low diffusion barriers [15-20]. $Mo_2C$ has been reported as a superconductor with a superconducting transition temperature of 9.7 K [21]. Generally, $Mo_2C$ adopts three crystalline structures [22-27]: orthorhombic, hexagonal, and trigonal. The ortho-$Mo_2C$ (O-MC) possesses a good thermal stability and is a good thermal barrier candidate in the multilayer stacks [28]. The hexagonal phase of $Mo_2C$ (H-MC) is well known to demonstrate excellent catalytic activity in a wide variety of reactions, such as hydrogenation of benzene [29], hydrogenation of alkanes [30], and alkane isomerization [31]. Furthermore, H-MC exhibits a strong-hydrogen binding energy.

To the best of our knowledge, a number of physical properties of O-MC, such as structural, elastic, electronic band structure, density of states, electron localization function, Fermi surfaces, phonon spectrum, heat capacity, Magnetic properties (Ni-doped), and superconducting state have been studied either experimentally or theoretically or so far [15, 32-41]. However, many other properties like Cauchy pressure, machinability index, mechanical anisotropy, acoustic ((both isotropic and anisotropic), acoustic impedance, electronic (charge density), both isotropic and anisotropic thermal conductivity (lattice, diffusion, and minimum), melting temperature, Mulliken population analysis, and optical properties are yet to be unveiled. Furthermore, there are only a few works on the H-MC system exist in the literature [22,32,42]. Quite surprisingly, most of the physical properties, e.g., elastic, electronic (DOS, charge density distribution, electron density difference, Fermi surface), phonon dispersion, thermophysical properties, Mulliken population analysis, and energy dependence of optical constants of the hexagonal phase of $Mo_2C$ have not been explored at all. All these unexplored properties have significant bearing on possible applications of H-MC. Therefore, a thorough understanding of the elastic, electronic, mechanical, acoustic, vibrational, electronic, thermal, bonding and optical constants spectra is important to unravel the potential of $Mo_2C$ for prospective applications. In this work we intend to bridge the existing research gap and explore the full potential of $Mo_2C$ in both the structures for applications in different sectors. We also plan to present a comparative analysis of the physical properties of $Mo_2C$ in both orthorhombic and hexagonal symmetries.

The rest of this manuscript is arranged as follows: Section 2 briefly describes the computational method. Section 3 presents and discusses the investigated properties with their possible implications. Finally, in Section 4 we summarize the key findings and draw conclusions of this work.

## 2. Computational methodology

The first-principles calculations were carried out using the density functional theory (DFT) [43,44] as contained within the Cambridge Serial Total Energy Package (CASTEP) code [45]. The exchange-correlation energy was evaluated using the generalized gradient approximation (GGA) within the Perdew-Burke-Ernzerhof for solids (PBEsol) [46] scheme. The plain Perdew-Burke-Ernzerhof (PBE) scheme [47] was found to overestimate equilibrium volume of $Mo_2C$ in both the structures. The interactions of valence electrons and the ion cores were modeled with the Vanderbilt-type ultrasoft pseudopotential [48]. The valence electron configurations for Mo and C elements of both the phases are taken as $4s^2$ $4p^6$ $4d^5$ $5s^1$ and $2s^2$ $2p^2$, respectively. Broyden–Fletcher–Goldfarb–Shanno (BFGS) minimization scheme [49] was applied for the geometry optimization of $Mo_2C$ which minimized the total energy and the internal forces. The plane-wave cutoff energy for both orthorhombic and hexagonal phases was set at 450 eV. This cut-off energy produced high level of convergence. The Brillouin zone (BZ) sampling was carried out using the Monkhorst and Pack scheme [50] with a mesh size of 7 × 6 × 7 and 18 ×18 × 11 for O-MC and H-MC symmetries, respectively. Various convergence tolerances during computations for both O-MC and H-MC phases were set to be less than $10^{-5}$ eV/atom for energy, maximum lattice point displacement within $10^{-3}$Å, maximum ionic force within 0.03 eVÅ$^{-1}$, maximum stress within 0.05 GPa and smearing width of 0.1 eV for electronic energy density of states, with finite basis set corrections [51]. For the Fermi surface construction denser k-point meshes of size, 24 × 20 × 24 were used for both O-MC and H-MC structures. All the present calculations of our compounds are performed at ground state with default temperature and pressure of 0 K and 0 GPa, respectively.

Elastic constants of O-MC and H-MC were calculated applying the stress–strain method [52]. Considering the crystal symmetry, an orthorhombic crystal has nine independent second-order elastic coefficients, which are $C_{11}$, $C_{22}$, $C_{33}$, $C_{44}$, $C_{55}$, $C_{66}$, $C_{23}$, $C_{12}$ and $C_{13}$. On the other hand, a hexagonal structure has six independent elastic constants, namely, $C_{11}$, $C_{33}$, $C_{44}$, $C_{66}$, $C_{12}$ and $C_{13}$. Those independent elastic constants $C_{ij}$ allowed us to evaluate all the macroscopic (polycrystalline) elastic moduli, such as the bulk modulus ($B$), shear modulus ($G$) and Young's modulus ($Y$) with the Voigte-Reuss-Hill (VRH) approximation, where applicable [53,54].

The frequency dependent optical spectra of both compounds are derived from the knowledge of complex dielectric function $\varepsilon(\omega) = \varepsilon_1(\omega) + i\varepsilon_2(\omega)$. The imaginary part of the dielectric function, $\varepsilon_2(\omega)$ has been evaluated from the CASTEP supported formula expressed as:

$$\varepsilon_2(\omega) = \frac{2e^2\pi}{\Omega\varepsilon_0} \sum_{k,v,c} |\langle \Psi_k^c | \hat{u}.\vec{r} | \Psi_k^v \rangle|^2 \; \delta(E_k^c - E_k^v - E) \quad (1)$$

In this equation, Ω refers the unit cell volume, ω is the frequency of incident photon, $\varepsilon_0$ is the permittivity of the free space, *e* is the electronic charge, ***u*** is defining the polarization of the incident electric field, ***r*** is the position vector, and $\Psi_k^c$ and $\Psi_k^v$ define the conduction and valence band wave functions at a given wave-vector *k*, respectively. This formula uses the inputs from the electronic band structure calculations. The real part of the dielectric function, $\varepsilon_1(\omega)$, has been evaluated from its corresponding imaginary part $\varepsilon_2(\omega)$ via the Kramers-Kronig transformation. Once the dielectric function is known, all the other optical constants, such as refractive index *n*(ω), absorption coefficient α(ω), energy loss-function *L*(ω), reflectivity *R*(ω), and optical conductivity *σ*(ω), can be deduced from it using standard formalism [55,56].

The projection of the plane-wave states onto a linear combination of atomic orbital (LCAO) basis sets [57,58] has been used for understanding the bonding nature in O-MC- and H-MC materials invoking Mulliken bond population analysis [59]. The Mulliken density operator written on the atomic (or quasi-atomic) basis has been used for the bond population analysis:

$$P_{\mu\nu}^{M}(g) = \sum_{g'}\sum_{\nu'} P_{\mu\nu'}(g')S_{\nu'\nu}(g-g') = L^{-1}\sum_{k} e^{-ikg}\,(P_k S_k)_{\mu\nu'} \qquad (2)$$

and the net charge on atom *A* is defined as

$$Q_A = Z_A - \sum_{\mu\in A} P_{\mu\mu}^{M}(0) \qquad (3)$$

where, $Z_A$ represents the charge on the atomic nucleus.

## 3. Results and analysis

### 3.1. Structural properties

The optimized crystal structures of $Mo_2C$ in orthorhombic and hexagonal phases are depicted in Figure 1. The orthorhombic phase of $Mo_2C$ is in *ξ*-$Fe_2N$-type crystal structure with space group *Pbcn* (No. 60). The unit cell of orthorhombic $Mo_2C$ includes twelve atoms with eight molybdenum and four carbon atoms. Every Mo atom possesses three C neighbors, while each C atom is coordinated with six Mo atoms. The carbon atoms in orthorhombic $Mo_2C$ regulate themselves in such way that each molybdenum atom possesses three almost planar carbon neighbors while each carbon atom coordinates with six molybdenum atoms within the structure. A two-dimensional (2D) view of $Mo_2C$ orthorhombic structure in the yz-plane shows the composition of hexagonal close-packed (HCP) molybdenum (Mo) atoms and interstitial carbon atoms orderly located in half Mo octahedral. On the other hand, $Mo_2C$ crystallizes in the hexagonal space group P63/mmc (No. 194). The unit cell consists of four atoms with two carbon atoms and two molybdenum atoms. The formula unit per unit cell is therefore, 1.33. The Wyckoff positions of Mo and C atoms in the orthorhombic and hexagonal phases of $Mo_2C$ are listed in Table 1. The results we got from the geometry optimization of O-MC and H-MC along with available experimental [22] and theoretical [32,42] data are listed in Table 2. It is evident that the present results are in good agreement with the previous experimental values. We can see that the equilibrium volume and the total number of atoms in the unit cell of hexagonal crystal are much smaller than those in the orthorhombic phase.

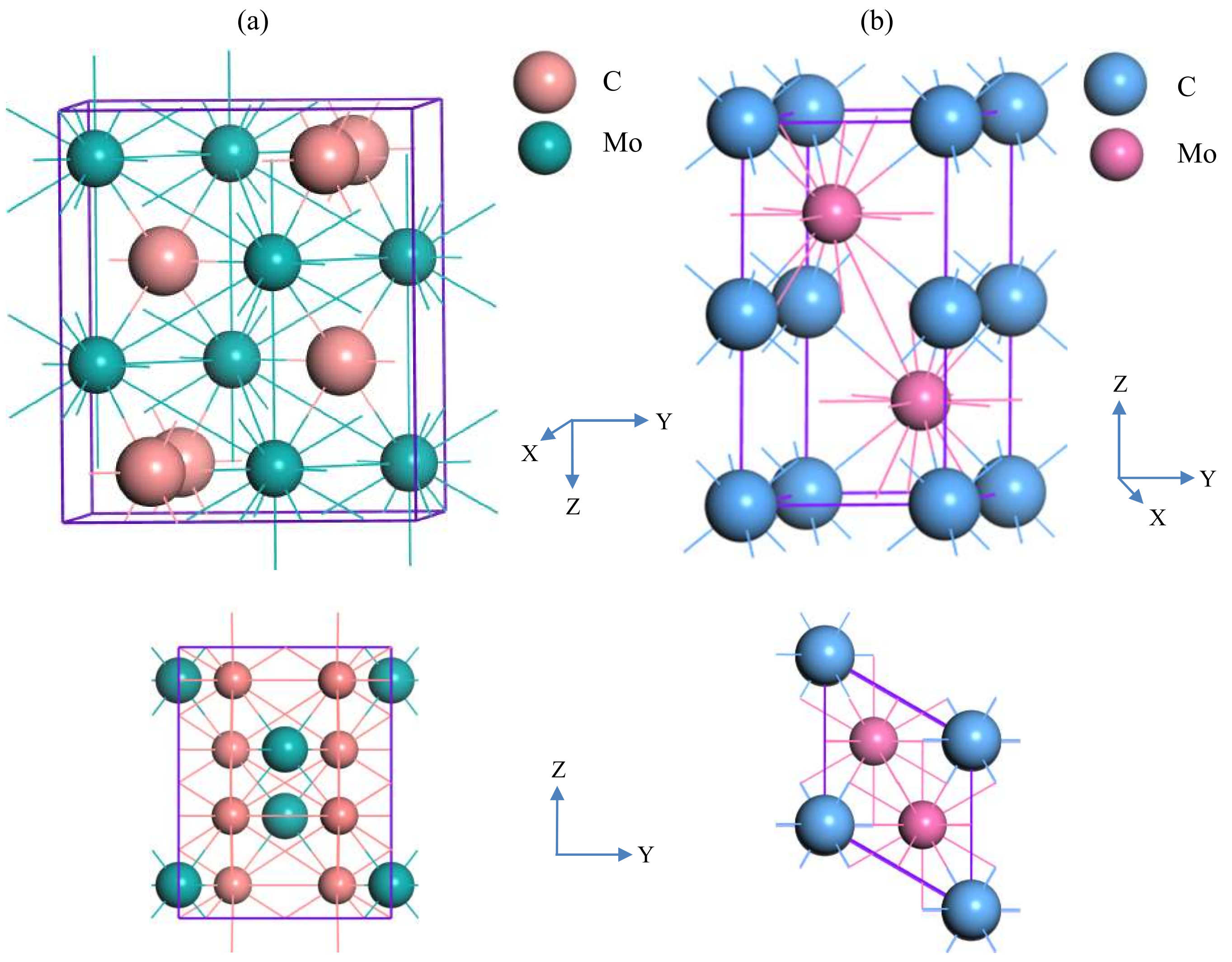

**Figure 1.** Schematic 3D Crystal structure and top view (2D view in YZ-plane) of (a) O-MC and (b) H-MC, respectively.

**Table 1.** Wyckoff positions of Mo and C atoms in the unit cell of O-MC and H-MC structures.

| | Mo-site | C-site | Remarks |
|---|---|---|---|
| O-MC | 0.250, 0.125, 0.083 | 0.500, 0.375, 0.250 | [22]$^{Expt.}$ |
| H-MC | 0.333, 0.667, 0.25 and 0.667, 0.333, 0.75 | 0, 0, 0 and 0, 0, 0.50 | |

**Table 2.** Calculated equilibrium lattice parameters (Å), equilibrium volume $V_0$ (Å$^3$), total number of atoms in the cell and formula unit ($Z$) of O-MC and H-MC structures together with previous results.

| Compounds | *a* | *b* | *c* | *c/a* | *b/a* | $V_0$ | No. of atoms | *Z* | Remarks |
|---|---|---|---|---|---|---|---|---|---|
| O-MC | 4.705 | 6.008 | 5.172 | 1.099 | | 146.235 | 12 | 4 | This work |
| | 4.725 | 6.022 | 5.195 | - | - | 147.46 | - | - | [22]$^{\text{Expt.}}$ |
| | 4.74 | 6.03 | 5.21 | - | - | - | - | - | [23]$^{\text{Expt.}}$ |
| | 4.738 | 6.038 | 5.210 | - | - | 149.09 | - | - | [32]$^{\text{Theo.}}$ |
| | 4.79 | 6.01 | 5.21 | - | - | - | - | - | [34]$^{\text{Theo.}}$ |
| | 4.75 | 6.09 | 5.26 | | - | | | | [33]$^{\text{Theo.}}$ |
| H-MC | 2.9260 | 2.9260 | 5.3729 | 1.836 | - | 39.8379 | 4 | 1.333 | This work |
| | 3.002 | 3.002 | 4.724 | - | - | - | - | - | [22]$^{\text{Expt.}}$ |
| | 6.004 | 6.004 | 4.724 | - | - | - | - | - | [42]$^{\text{Expt.}}$ |
| | 6.098 | 6.098 | 6.098 | - | - | 150.17 | - | - | [32]$^{\text{Theo.}}$ |

### 3.2. Mechanical properties

It is important to characterize the mechanical behavior of a material in order to understand its practical application limits under externally applied mechanical loads different types. The single crystal elastic constants are known as fundamental and indispensable parameters for describing the mechanical properties of materials. It provides a direct link between the mechanical properties and dynamic information concerning the nature of the forces operating in solids, especially for the stability and stiffness of materials. The mechanical properties such as stability under static and time-varying stress, plasticity, stiffness, brittleness, ductility, and elastic anisotropy of compounds are closely associated with the elastic constants. The independent elastic constants for O-MC and H-MC are presented in Table 3 together with prior results. According to Born-Huang criteria, mechanical stability conditions different for different crystal structures. The inequality criteria for orthorhombic crystals [60,61] are: $C_{11} > 0$, $C_{22} > 0$, $C_{33} > 0$, $C_{44} > 0$, $C_{55} > 0$, $C_{66} > 0$, $(C_{11} + C_{22} - 2C_{12}) > 0$, $(C_{11} + C_{33} - 2C_{13}) > 0$, $(C_{22} + C_{33} - 2C_{23}) > 0$, $(C_{11} + C_{22} + C_{33} + 2C_{12} + 2C_{13} + 2C_{23}) > 0$; and for hexagonal systems [62] we have: $C_{11} - |C_{12}| > 0$, $(C_{11} + C_{12})C_{33} - 2{C_{13}}^2 > 0$, $C_{44} > 0$. The computed values of elastic constants meet the above stability criteria, indicating that both O-MC and H-MC are mechanically stable crystalline systems. The resistance to linear compression and atomic bonding along [100], [010] (where applicable) and [001] directions are directly related to $C_{11}$, $C_{22}$ and $C_{33}$, respectively. For both the structures, the value of $C_{33}$ is considerably larger than $C_{11}$, indicating that atoms along [001] are strongly bonded as compared to those along [100] direction and the structures are more compressible when stress is along [100] direction. Another elastic constant, $C_{44}$, is usually used for measuring the resistance to shear deformation for tangential stress applied to the (100) plane in the [010] direction of the compound [63]. $C_{44}$ is much higher for H-MC than that for O-MC, implying a higher shear deformation resistance for hexagonal phase compared to the orthorhombic phase. Also, $C_{44}$ is much lower than $C_{11}$ and $C_{33}$ for both phases, which indicates that both compounds are more easily deformed by a shear in comparison to a unidirectional stress along any of the three principal crystallographic directions. $C_{44}$ also represents a resistance to shape change without volume change, and provides with information about electronic response to shear strain [64]. In addition, the higher values of $C_{11}$ and $C_{33}$ than $C_{12}$, $C_{13}$, and $C_{14}$, implying that shear deformation is easier than axial deformation for both phases. In both cases, the lower value of $C_{44}$ compare to $C_{66}$ predicts that the shear in the (100) plane is easier relative to the shear in the (001) plane.

The calculated elastic constants lead to the calculation of the polycrystalline elastic moduli via the Voigt-Reuss-Hill (VRH) approximations [65-67]. The Voigt and Reuss approximations yield extreme values; the Voigt (V) approximation provides the upper limits of $B$ and $G$, whereas the Reuss (R) approximation defines the lower limits. Moreover, Hill (H) recommended that the effective bulk and shear moduli be estimated by an arithmetic mean of the Voigt and Reuss limits. The bulk, shear, and Young moduli of O-MC and H-MC are calculated from the following standard relations:

$$B_V = \frac{[C_{11} + C_{22} + C_{33} + 2(C_{12} + C_{13} + C_{23})]}{9} \tag{4}$$

$$G_V = \frac{[C_{11} + C_{22} + C_{33} + 3(C_{44} + C_{55} + C_{66}) - (C_{12} + C_{13} + C_{23})]}{15} \tag{5}$$

$$B_H = \frac{B_V + B_R}{2};\ G_H = \frac{G_V + G_R}{2} \tag{6}$$

$$E_{VRH} = 9B_{VRH}G_{VRH}/(3B_{VRH} + G_{VRH}) \tag{7}$$

The obtained values are listed in Table 4. The bulk modulus of O-MC and H-MC are quite high. Both orthogonal and hexagonal phases are highly incompressible with $B$ of 314.31 and 368.75 GPa, respectively. These values are comparable to the hardest and the least compressible material known so far; diamond (with Bulk modulus 442 GPa) [68]. The bulk and shear modulus of a solid represent the rigidity against fracture and plastic deformation, respectively. The mechanical strength of both compounds will be limited by the shear deformation since $G$ is smaller than $B$ (Table 4) in both cases. O-MC structure exhibits stronger volume and shear deformation resistances as well as higher stiffness compared to H-MC. As a general rule the bulk modulus of a solid is inversely related to its unit cell volume by $B \sim V^k$ [69]. The fracture strength of all pure metals is proportional to $B$. The larger the bulk modulus, the larger the average atomic bond strength of materials because of its strong correlation with the cohesive energy or binding energy of atoms in a molecule [70]. On the other hand, large value of shear modulus is an indicator of pronounced directional bonding between the atoms [71]. The Pugh's ratio, *G/B*, is a widely used parameter to demarcate brittleness from ductility [72]. The critical value used to separates ductile and brittle materials is around 0.57 (for brittle materials, $G/B > 0.57$, and for ductile materials, $G/B < 0.57$). The *G*/*B* values of O-MC and H-MC are 0.51 and 0.54, respectively. The calculated values show that both structures are ductile in nature. Tanaka *et al*. [73] proposed that *G/B* also represents the relative directionality of the bonding in the material considering bulk and shear modulus as a measure of the average bond strength and the resistance to a change in bond angle by an external force, respectively. The value of *G/B* for H-MC is larger than that for O-MC, indicating that the directionality of the atomic bonding in H-MC is stronger. Young modulus (*E*) measures the stiffness of solids and quantifies the resistance to longitudinal stress. In contrast to the situation regarding bulk modulus, the Youngs modulus is higher for the H-MC structure as seen in Table 4. The Poisson's ratio (υ) of a material can be calculated using *B* and *G* from following expression [63]:

$$\nu = \frac{(3B - 2G)}{2(3B + G)} \tag{8}$$

Table 4 shows that H-MC is predicted to be stiffer than O-MC. Poisson's ratio ($\nu$) is a widely used indicator for compressibility, brittleness/ductility, and bonding nature of a solid. It is intimately connected with the way structural elements are packed. Materials with different Poisson's ratios behave very differently under external stress. The value of $\nu$ of all stable materials ranges between -1.0 to 0.5. The volume of materials remains unchanged with any amount of deformation when ν is equal to 0.5. These materials are called incompressible (viscoelastic) solids. Solids with $\nu > 0.26$ and $\nu < 0.26$ are classified as ductile and brittle, respectively [74]. The presence of central interatomic forces (acting along the lines joining pairs of atoms) in materials can be predicted from Poisson ratio. The lower and upper limits of $\nu$ for central force are 0.25 and 0.50, respectively [75,76]. The typical values of $\nu$ for purely covalent and ionic materials are of the order of 0.10 and 0.25, respectively [77]. The estimated values of $\nu$ of O-MC and H-MC are 0.28 and 0.27, respectively (Table 4). Thus, we can predict that both the structures of $Mo_2C$ are ductile, significantly ionic bonded and the interatomic forces are central in nature. Besides, the relationship between bulk and shear moduli for covalent and ionic materials are: $G \sim 1.1\,B$ and $G \sim 0.6\,B$, respectively [78].

Another fundamental mechanical parameter for material characterization is called Cauchy pressure, $P_C$. The expression of $P_C$ is different for different symmetries [79-81].

$$P_C^a = C_{23} - C_{44}\,, P_C^b = C_{13} - C_{55}, P_C^c = C_{12} - C_{66}\ (Orthorhombic\ symmetry) \tag{9}$$

$$P_C^a = C_{13} - C_{44}\,, P_C^b = C_{12} - C_{66}\ (Hexagonal\ symmetry) \tag{10}$$

For hexagonal crystals, the Cauchy pressures for (100) and (001) planes are defined as ($C_{13} - C_{44}$) and ($C_{12} - C_{66}$), respectively. It reflects the nature of the bonding in materials at the atomistic level. For metallic-like bonding, $P_C$ must be positive. The Cauchy pressure is another approach to describe the brittleness/ductility of materials [82]. If the Cauchy pressure is positive, then the material is ductile, whereas negative Cauchy pressure demonstrates brittleness. Table 5 shows the calculated Cauchy pressures of O-MC and H-MC. The positive Cauchy pressures for the compounds clearly indicate that both of them are ductile in nature. Cauchy pressure can be used to describe the presence of angular characteristics of atomic bonding in solids. According to Pettifor [79], positive Cauchy pressure reflects metallic non-directional bonding, whereas the negative Cauchy pressure is for nonmetallic materials with angular character of bonding. The more negative the Cauchy pressure, more directional characteristic, lesser bonding mobility, and more brittleness, the material exhibits. If the bonding can be explained by simple pair-wise potentials, then Cauchy pressure will be zero. The positive values of $P_C$ for both structures suggest ductile nature with presence of delocalized metallic bondings. The predictions from Cauchy pressure are in complete accord with those associated with the Paugh's ratio and Poisson's ratio predictions. For O-MC, the order of Cauchy pressure is $P_C^b > P_C^c > P_C^a$. For H-MC, the Cauchy pressure ($C_{12} - C_{66}$) is larger than the counterpart ($C_{13} - C_{44}$), implying that the metallic character of the bonding in the (001) plane is more significant than that in the (100) plane.

Hard materials with high level of machinability are of great interest in engineering applications. Machinability explains the ability of a material to be machined using machine tools. It controls cutting (speed, force), cutting energy, drilling rates, feed rate, tool ware, machining time, and depth of cut. The machinability depends on many other external factors, such as work material, cutting tool used, and the cutting parameters. The plasticity [83-86] and lubricating property of materials also correlate with machinability. The higher value of plastic deformation indicates better machinability index. The machinability index, $\mu_M$ of a material can be defined as [87]:

$$\mu_M = \frac{B}{C_{44}} \tag{11}$$

Therefore, high modulus of elasticity with low shear resistance leads to good machinability and better dry lubricity. The thermal conductivity of a solid has influence on limiting its machinability. Solids with good machinability exhibit excellent lubricating properties, lower feed forces, lower friction value, and higher plastic strain value. The calculated values of $\mu_M$ are disclosed in Table 4. The machinability index of orthorhombic structure is larger than that in the hexagonal structure. Hence, O-MC should be more machinable than the H-MC.

Ultra-incompressible and superhard materials are of significant importance in industrial applications due to their outstanding properties such as high elastic modulus and hardness, scratch resistance, surface durability as well as chemical stability. Hard materials are used for cutting tools and wear-resistant coatings. In this study, the following semi-empirical correlations [88-90] between Vickers hardness ($H_V$) and $B$, $G$, $E$, $v$, and $B/G$, the macroscopic models for hardness prediction, are used to calculate hardness of O-MC and H-MC materials:

$$H_1 = 0.0963B \tag{12}$$

$$H_2 = 0.0607E \tag{13}$$

$$H_3 = 0.1475G \tag{14}$$

$$H_4 = 0.0635E \tag{15}$$

$$H_5 = -2.899 + 0.1769G \tag{16}$$

$$H_6 = \frac{(1-2\upsilon)B}{6(1+\upsilon)} \tag{17}$$

$$H_7 = \frac{(1-2\upsilon)E}{6(1+\upsilon)} \tag{18}$$

$$H_8 = 2(k^2G)^{0.585} - 3 \tag{19}$$

Here, $k$ = $G/B$ (Pugh's ratio), $B$ and $G$ are in GPa. The aptitude of the above methods in predicting hardness depends of crystal class and energy band gap ($E_g$) in case of semiconductors [91]. To select the best hardness estimators for orthorhombic and hexagonal crystal structures, we have calculated hardness of O-MC and H-MC, displayed in Table 6, using all these formulas. The best model for hardness calculations of orthorhombic and hexagonal crystals among these seven hardness analysis methods is $H_8$ [91]. Both structures show high value of hardness and H-MC is predicted to be harder than O-C. This agrees well with the $C_{44}$ values of the two structures. This parameter is a useful to roughly predict hardness of solids.

**Table 3.** Calculated elastic constants ($C_{ij}$) (in GPa) of O-MC and H-MC crystals of $Mo_2C$.

| | $C_{11}$ | $C_{22}$ | $C_{23}$ | $C_{33}$ | $C_{12}$ | $C_{44}$ | $C_{13}$ | $C_{55}$ | $C_{66}$ | Remarks |
|---|---|---|---|---|---|---|---|---|---|---|
| O-MC | 485.839 | 515.322 | 182.951 | 518.370 | 236.612 | 155.274 | 235.418 | 178.659 | 197.625 | This work |
| | 460.04 | 495.17 | 185.77 | 492.45 | 228.67 | 140.61 | 238.62 | 161.81 | 183.25 | [33][Theo.] |
| | 430.3 | 494.4 | 164.4 | 485.6 | 214.9 | 145.5 | 217.6 | 161.7 | 178.2 | [34][Theo.] |
| | 460.5 | 482.7 | 166.9 | 489.4 | 210.9 | 148.2 | 205.8 | 174.7 | 182.4 | [32][Theo.] |
| | 421 | 489 | 137 | 491 | 240 | 143 | 227 | 180 | 172 | [38][Theo.] |
| | 454 | 487 | 166 | 490 | 224 | 146 | 212 | 184 | 171 | [90][Theo.] |
| | 453 | 489 | 169 | 488 | 216 | 139 | 226 | 167 | 179 | [91][Theo.] |
| H-MC | 628.493 | - | - | 710.973 | 235.804 | 189.26 | 220.593 | - | 196.345 | This work |

**Table 4.** The calculated isotropic bulk modulus $B$ (GPa), shear modulus $G$ (GPa), Young's modulus $E$ (GPa), Pugh's indicator $G/B$, Machinability index $\mu_M$, Poisson's ratio ν, and hardness H (GPa) of O-MC and H-MC (Hexa) compounds deduced from Voigt-Reuss-Hill (VRH) approximations.

| | $B$ | | | $G$ | | | $E$ | $\frac{B_V}{B_R}$ | $\frac{G_V}{G_R}$ | $G/B$ | $\mu_M$ | $v$ | Remarks |
|---|---|---|---|---|---|---|---|---|---|---|---|---|---|
| | $B_V$ | $B_R$ | $B_H$ | $G_V$ | $G_R$ | $G_H$ | | | | | | | |
| O-MC | 314.39 | 314.24 | 314.31 | 163.95 | 159.54 | 161.74 | 414.19 | 1.0004 | 1.028 | 0.51 | 2.02 | 0.28 | This work |
| | - | - | 305.9 | - | - | 147.9 | 382.3 | - | - | - | - | 0.29 | [33]Theo. |
| | - | - | 289.3 | - | - | 149.1 | 381.7 | - | - | - | - | 0.28 | [34]Theo. |
| H-MC | 369.10 | 368.39 | 368.75 | 201.04 | 199.72 | 200.38 | 508.95 | 1.002 | 1.007 | 0.54 | 1.95 | 0.27 | This work |

**Table 5.** Estimated Cauchy pressure ($P_C$) of O-MC and H-MC.

| Compounds | $P_C^a$ | $P_C^b$ | $P_C^c$ |
|---|---|---|---|
| O-MC | 27.677 | 56.759 | 38.987 |
| H-MC | 31.333 | 39.459 | - |

**Table 6.** The calculated values of hardness of O-MC and H-MC from different empirical formulas.

| Compounds | $H_1$ | $H_2$ | $H_3$ | $H_4$ | $H_5$ | $H_6$ | $H_7$ | $H_8$ | Remarks |
|---|---|---|---|---|---|---|---|---|---|
| O-MC | 30.27 | 25.14 | 23.86 | 26.30 | 25.71 | 18.01 | 23.73 | 14.82 | This work |
| | - | - | - | - | - | - | - | 14.21 | [34]Theo. |
| | - | - | - | - | | - | - | 14 | [92]Expt. |
| H-MC | 35.51 | 30.89 | 29.56 | 32.32 | 32.55 | 22.26 | 30.72 | 18.60 | This work |

### 3.3. Elastic anisotropy

Mechanical anisotropy, one of the key features of crystalline solids, has important implications in engineering science. The creation and propagation of micro-cracks in materials, direction dependence of chemical bonding, and mechanical durability of materials are directly related to elastic anisotropy. Generally, directional covalent bonding contributes to the crystal's anisotropy, whereas metallic bonding improves overall isotropy [93,94].

The degree of anisotropy in atomic bonding in different crystallographic planes can be explained from shear anisotropy factor. The shear anisotropy for different planes of the orthorhombic and hexagonal crystals are defined below [71,95]:

The shear anisotropic factor for {100} shear planes between the ⟨011⟩ and ⟨010⟩ directions is,

$$A_1 = \frac{4C_{44}}{C_{11} + C_{33} - 2C_{13}} \tag{20}$$

for the {010} shear plane between ⟨101⟩ and ⟨001⟩ directions is,

$$A_2 = \frac{4C_{55}}{C_{22} + C_{33} - 2C_{23}} \tag{21}$$

and for the {001} shear planes between ⟨110⟩ and ⟨010⟩ directions is,

$$A_3 = \frac{4C_{66}}{C_{11} + C_{22} - 2C_{12}} \tag{22}$$

The calculated shear anisotropic factors of O-MC and H-MC are listed in Table 7. All three factors must be equal to unity in the case of isotropy in shearing response. The deviation from unity is a measure of degree of anisotropy. For orthorhombic structure, the computed values of $A_1$, $A_2$ and $A_3$ are 1.16, 1.07 and 1.50, respectively. For hexagonal structure $A_1 = A_2 = 0.84$ and $A_3 = 1$. Thus, the H-MC shows shear isotropy for the {001} plane. The estimated values predict that both compounds are moderately anisotropic but O-MC exhibits more anisotropy than the H-MC one.

The universal anisotropy index $A^U$, equivalent Zener anisotropy measure $A^{eq}$, anisotropy in shear $A^G$ (or $A^C$) and anisotropy in compressibility $A^B$ of solids with any symmetry can be estimated from the following standard formulas [95-98]:

$$A^U = 5\frac{G_V}{G_R} + \frac{B_V}{B_R} - 6 \geq 0 \tag{23}$$

$$A^{eq} = \left(1 + \frac{5}{12}A^U\right) + \sqrt{\left(1 + \frac{5}{12}A^U\right)^2 - 1} \tag{24}$$

$$A^G = \frac{G^V - G^R}{2G^H} \tag{25}$$

$$A^B = \frac{B_V - B_R}{B_V + B_R} \tag{26}$$

The computed results of these parameters are also listed in Table 7. Universal elastic anisotropic index ($A^U$) is one of the most widely used indices to qualify the anisotropy because of its applicability to all possible crystal symmetries. $A^U$ is the first anisotropy parameter among all other anisotropy measures which accounts both shear and bulk contribution. It is obvious from Eqn. 23 that $G_V/G_R$ has greater influence on the anisotropy index $A^U$ than $B_V/B_R$. $A^U$ is zero for the case of locally isotropic single crystals, whereas the degree of deviation from zero suggests varying extent of anisotropy in materials. The calculated values of $A^U$ for O-MC and H-MC are 0.14 and 0.04, respectively. Thus, according to this particular indicator, the anisotropy in O-MC is higher than that in H-MC.

For locally isotropic solids, $A^{eq}$ is equal to 1.0. The calculated values of $A^{eq}$ for O-MC and H-MC are 1.40 and 1.19, respectively. These values predict moderate anisotropy of the studied systems. For both $A^B$ and $A^G$, values of 0.0 and 1.0 (100%) represent elastic isotropy and the largest possible anisotropy, respectively. The larger value of $A^G$ in comparison with $A^B$ (Table 7) for both structures indicates that anisotropy in shear is higher than the anisotropy in compressibility.

The universal log-Euclidean index can be defined using the log-Euclidean formula as [96,99]:

$$A^L = \sqrt{\left[ln\left(\frac{B^V}{B^R}\right)\right]^2 + 5\left[ln\left(\frac{C_{44}^V}{C_{44}^R}\right)\right]^2} \tag{27}$$

Here, $C_{44}^V$ and $C_{44}^R$ refer to the Voigt and Reuss values of $C_{44}$, respectively.

The values of $C_{44}^V$ and $C_{44}^R$ are obtained from [96]:

$$C_{44}^R = \frac{5}{3}\frac{C_{44}\,(C_{11} - C_{12})}{3(C_{11} - C_{12}) + 4C_{44}} \tag{28}$$

and

$$C_{44}^V = C_{44}^R + \frac{3}{5}\frac{(C_{11} - C_{12} - 2C_{44})^2}{3(C_{11} - C_{12}) + 4C_{44}} \tag{29}$$

This index is correctly scaled for perfect isotropy and is valid also for all the crystalline symmetries. However, $A^L$ is an absolute measure of anisotropy in crystalline materials since it is less sparse than $A^U$ when considering extremely anisotropic crystals. The drawback of $A^U$ is that it only explains presence of anisotropy in a material, not the absolute level of anisotropy. So, $A^L$ is considered to be a more appropriate parameter for anisotropy study. The values of $A^L$ range between 0 to 10.27, and around 90% of solids having an anisotropy $A^L < 1.0$ [96]. For a perfectly isotropic material, $A^L$ is equal to zero. The larger the $A^L$, the higher the anisotropy. The estimated values of $A^L$ for O-MC and H-MC are 0.077 and 0.003, respectively. It is obvious that both are mildly anisotropic, but O-MC shows higher level of anisotropy. Generally, it has been claimed that materials with higher (lower) $A^L$ possesses layered (non-layered) type structure [96,100,101]. The comparatively low value of $A^L$ implies that both studied compounds exhibit non-layered structural features.

The linear compressibility of solids along $a$ and $c$ directions ($\beta_a$ and $\beta_c$) can be evaluated from [102]:

$$\beta_a = \frac{C_{33} - C_{13}}{D} \quad \text{and} \quad \beta_c = \frac{C_{11} + C_{12} - 2C_{13}}{D} \tag{30}$$

with, $D = (C_{11} + C_{12})C_{33} - 2(C_{13})^2$

The calculated values of these parameters are listed in Table 7. The ratio of linear compressibility along the $c$-axis to that along the $a$-axis, $\beta_c/\beta_a$, have unit value for isotropic solids. The deviation of these factors from unity quantifies the level of anisotropic compressibility. The estimated values indicate that the linear compressibility of both compounds exhibits considerable anisotropy. O-MC possesses higher anisotropy in compressibility than H-MC. Also, the compressibility along $a$-axis is more than that along the $c$-axis for both compounds.

**Table 7.** Shear anisotropy factor ($A_1$, $A_2$ and $A_3$), the universal anisotropy index $A^U$, equivalent zener anisotropy measure $A^{eq}$, anisotropy in shear $A_G$ (or $A^C$), anisotropy in compressibility $A_B$, universal log-Euclidean index $A^L$, linear compressibility ($\beta_a$ and $\beta_c$) ($TPa^{-1}$) and their ratio $\beta_c/\beta_a$ for O-MC and H-MC in the ground state.

| Compounds | $A_1$ | $A_2$ | $A_3$ | $A^U$ | $A^{eq}$ | $A_G$ | $A_B$ ($\times10^{-4}$) | $A^L$ | *Layered* | $\beta_a$ ($\times10^{-3}$) | $\beta_c$ ($\times10^{-3}$) | $\beta_c/\beta_a$ | Remarks |
|---|---|---|---|---|---|---|---|---|---|---|---|---|---|
| O-MC | 1.16 | 1.07 | 1.50 | 0.14 | 1.40 | 0.014 | 2.37 | 0.077 | No | 1.07 | 0.95 | 0.889 | This work |
| | - | - | - | 0.15 | - | 0.014 | 1 | - | - | - | - | - | [33]Theo. |
| | - | - | - | 0.15 | - | 0.015 | 0.001 | - | - | - | - | - | [34]Theo. |
| H-MC | 0.84 | 0.84 | 1 | 0.04 | 1.19 | 0.003 | 9.71 | 0.003 | No | 0.95 | 0.82 | 0.863 | This work |

Unlike in a cubic crystal, the bulk modulus of a crystal has directional dependence and the elastic anisotropy arises from the anisotropy of the linear bulk modulus in addition to the shear anisotropy. Hence, the shear anisotropy factors are not sufficient to describe elastic anisotropy. The bulk modulus of solids along the crystallographic axes can be calculated either from the pressure dependent lattice parameter or by means of the single crystal elastic constants. It is easy to calculate the bulk modulus along three axes using the single crystal elastic constants. The relaxed bulk modulus and uniaxial bulk modulus along $a$, $b$ and $c$ axis and anisotropies of the bulk modulus of $Mo_2C$ are evaluated from following formulas [103]:

$$B_{relax} = \frac{\Lambda}{(1 + \alpha + \beta)^2} \; ; \; B_a = a\frac{dP}{da} = \frac{\Lambda}{1 + \alpha + \beta} \; ; \; B_b = a\frac{dP}{db} = \frac{B_a}{\alpha} \; ; \; B_c = c\frac{dP}{dc} = \frac{B_a}{\beta} \tag{31}$$

and

$$A_{B_a} = \frac{B_a}{B_b} = \alpha \quad ; \quad A_{B_c} = \frac{B_c}{B_b} = \frac{\alpha}{\beta} \tag{32}$$

where,

$$\Lambda = C_{11} + 2C_{12}\alpha + C_{22}\alpha^2 + 2C_{13}\beta + C_{33}\beta^2 + 2C_{23}\alpha\beta$$

$$\alpha = \frac{(C_{11} - C_{12})(C_{33} - C_{13}) - (C_{23} - C_{13})(C_{11} - C_{13})}{(C_{33} - C_{13})(C_{22} - C_{12}) - (C_{13} - C_{23})(C_{12} - C_{23})}$$

and

$$\beta = \frac{(C_{22} - C_{12})(C_{11} - C_{13}) - (C_{11} - C_{12})(C_{23} - C_{12})}{(C_{22} - C_{12})(C_{33} - C_{13}) - (C_{12} - C_{23})(C_{13} - C_{23})}$$

where, $A_{B_a}$ and $A_{B_c}$ represents anisotropies of bulk modulus along the $a$-axis and $c$-axis with respect to $b$-axis, respectively.

The calculated results are disclosed in Table 8. For orthorhombic structure, the bulk modulus along the $a$-axis is larger than those along the $b$- and $c$-axis. On the other hand, the highest directional bulk modulus is along the $c$-axis. Therefore, for orthorhombic and hexagonal compound, the compressibility along $a$-axis and $c$-axis is the smallest, respectively. The hexagonal structure possesses a larger bulk modulus in each direction than the orthorhombic structure. Generally, the uniaxial bulk modulus of a material is different and much larger from the isotropic bulk modulus. The reason being the pressure in a state of uniaxial strain for a given crystal density generally differs from the pressure in a state of hydrostatic stress at the same density of the solid [71]. A value of $A_{B_a} = A_{B_b} = 1.0$ implies elastic isotropy and any departure from unity represent elastic anisotropy.

**Table 8.** Anisotropies in bulk modulus along different axes of O-MC and H-MC compounds.

| Compounds | $B_{relax}$ | $B_a$ | $B_b$ | $B_c$ | $A_{B_a}$ | $A_{B_c}$ |
|---|---|---|---|---|---|---|
| O-MC | 314.24 | 1003.60 | 912.29 | 917.65 | 1.10 | 1.006 |
| H-MC | 368.39 | 1054.63 | 1054.63 | 1222.30 | 1.00 | 1.160 |

The 2D and 3D graphical presentations of Young modulus ($E$), linear compressibility ($\beta$), Shear modulus ($G$), and Poisson ratio (ν) of materials are useful to illustrate the elastic anisotropy of crystals. The elastic stiffness matrixes for both phases are calculated using CASTEP. This matrix is feed to the ELATE program [104] in order to direct visualization of anisotropy level in Young's modulus ($E$), linear compressibility ($\beta$), shear modulus ($G$), and Poisson's ratio ($v$) of $Mo_2C$. The uniform circular (2D) and spherical (3D) graphical representations are indicators of the isotropic nature of crystals. The greater the deviation from these shapes, the higher the level of anisotropy. Figure 2 and Figure 3 show 2D [in (xy)*ab*-, (xz)*ac*- and (yz)*bc*-planes] and 3D view of direction dependency of E, $\beta$, $G$ and $v$ for orthorhombic and hexagonal phases of $Mo_2C$, respectively. The green and blue curves show the minimum and the maximum points for the parameters, respectively. The negative values are shown by red curves. For both crystal structures, linear compressibility is almost isotropic in all the planes. In the case of hexagonal structure, all four parameters are isotropic in the *ab*-plane, while they are anisotropic in other planes. Both 2D and 3D graphical plots conclude that anisotropy increases following the order $\beta < E < G < v$ in both phases. Also, the maximum and minimum limits of these parameters along with their ratios are listed in Table 9.

**Table 9.** The minimum and maximum limits of Young's modulus (GPa), linear compressibility ($TPa^{-1}$), shear modulus (GPa), Poisson's ratio and their ratios for O-MC and H-MC.

| Compounds | $E$ | | $A_E$ | $\beta$ | | $A_\beta$ | $G$ | | $A_G$ | $v$ | | $A_v$ |
|---|---|---|---|---|---|---|---|---|---|---|---|---|
| | $E_{min}$ | $E_{max}$ | | $\beta_{min}$ | $\beta_{max}$ | | $G_{min}$ | $G_{max}$ | | $v_{min}$ | $v_{max}$ | |
| O-MoC | 326.63 | 469.46 | 1.437 | 0.997 | 1.096 | 1.100 | 128.17 | 197.62 | 1.542 | 0.174 | 0.401 | 2.301 |
| H-MoC | 489.76 | 598.37 | 1.222 | 0.818 | 0.948 | 1.159 | 189.26 | 222.97 | 1.178 | 0.218 | 0.313 | 1.437 |

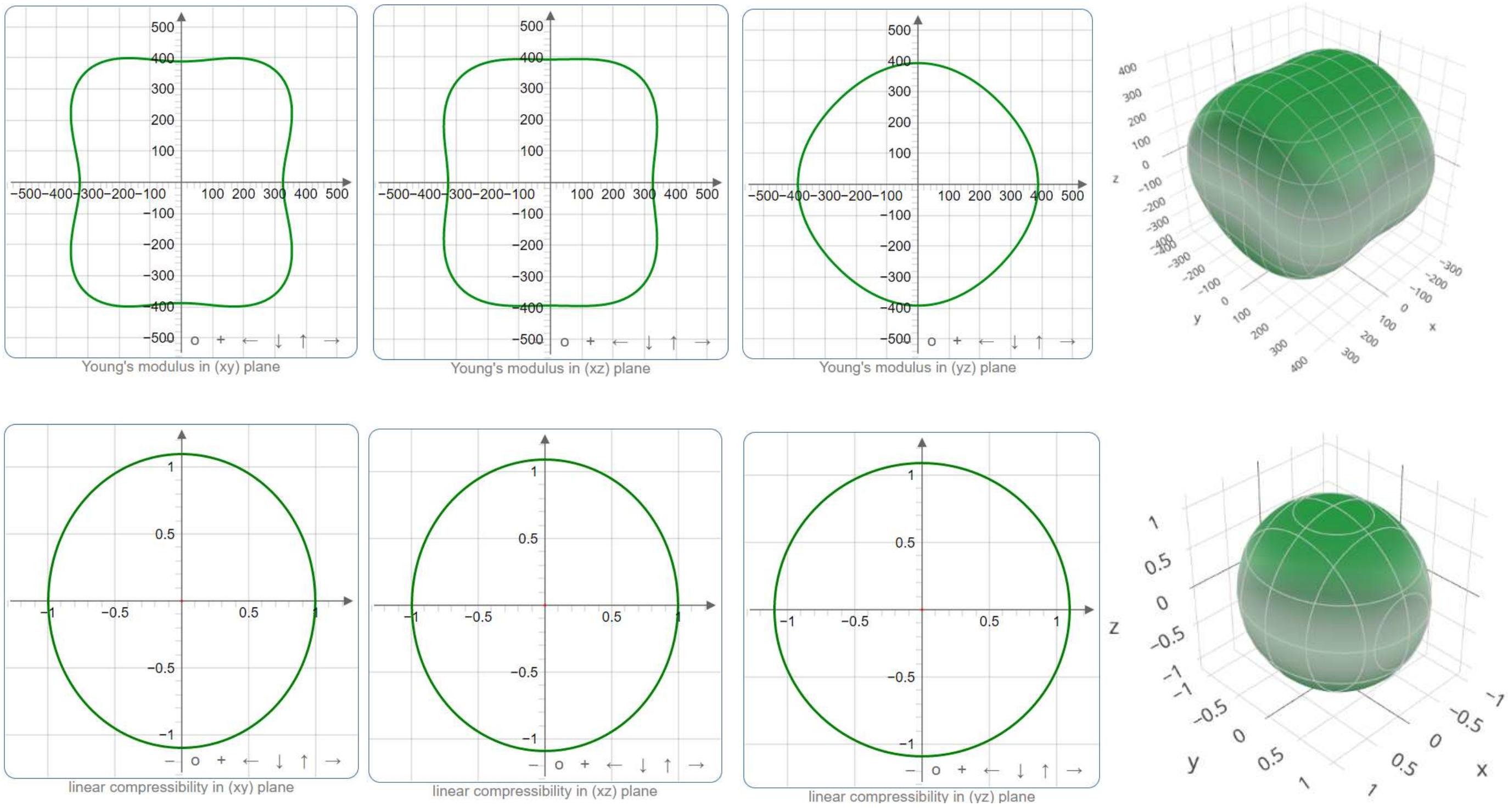

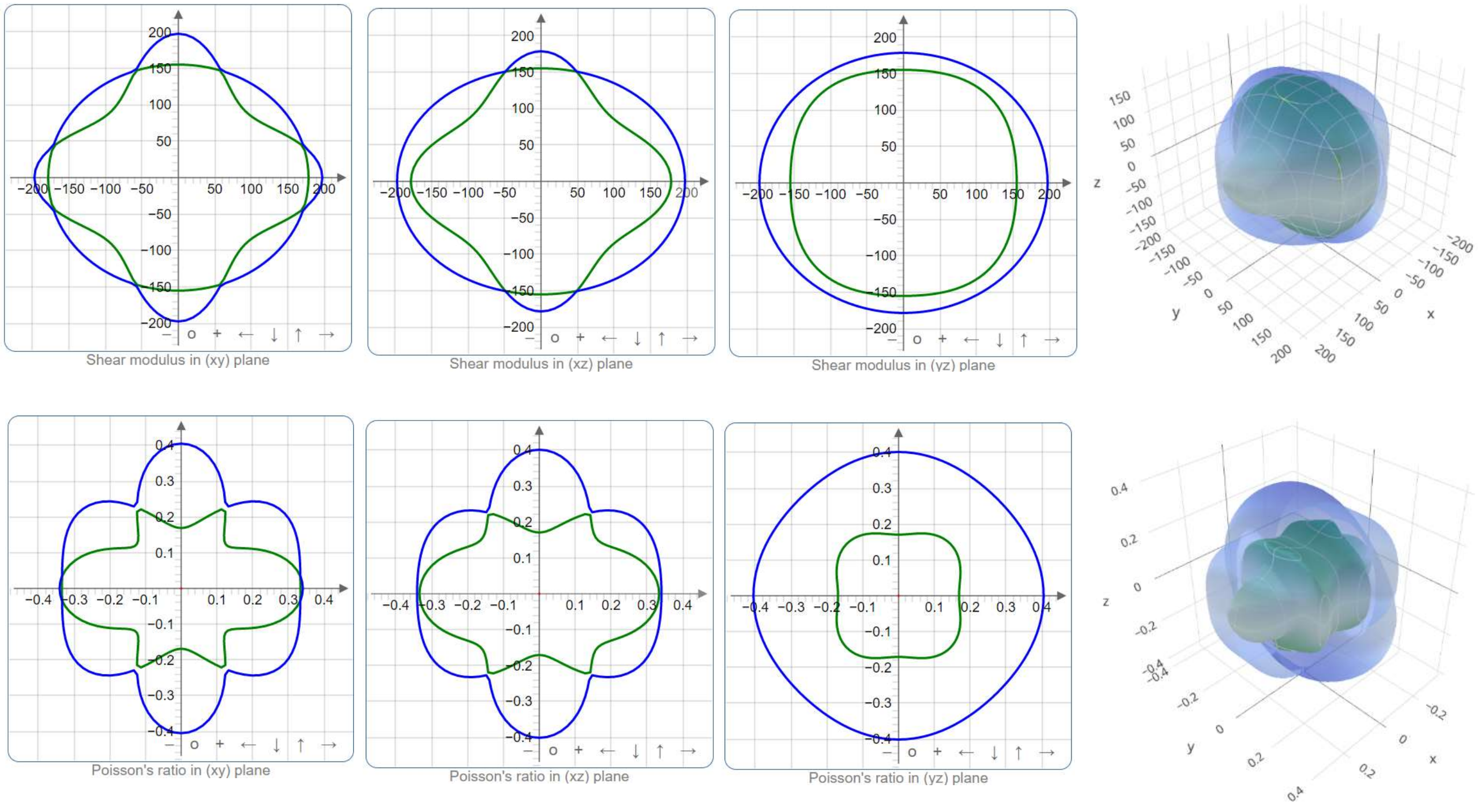

**Figure 2.** Directional dependence of Young's modulus (*E*), linear compressibility (*β*), shear modulus (*G*) and Poisson's ratio (ν) of O-MC.

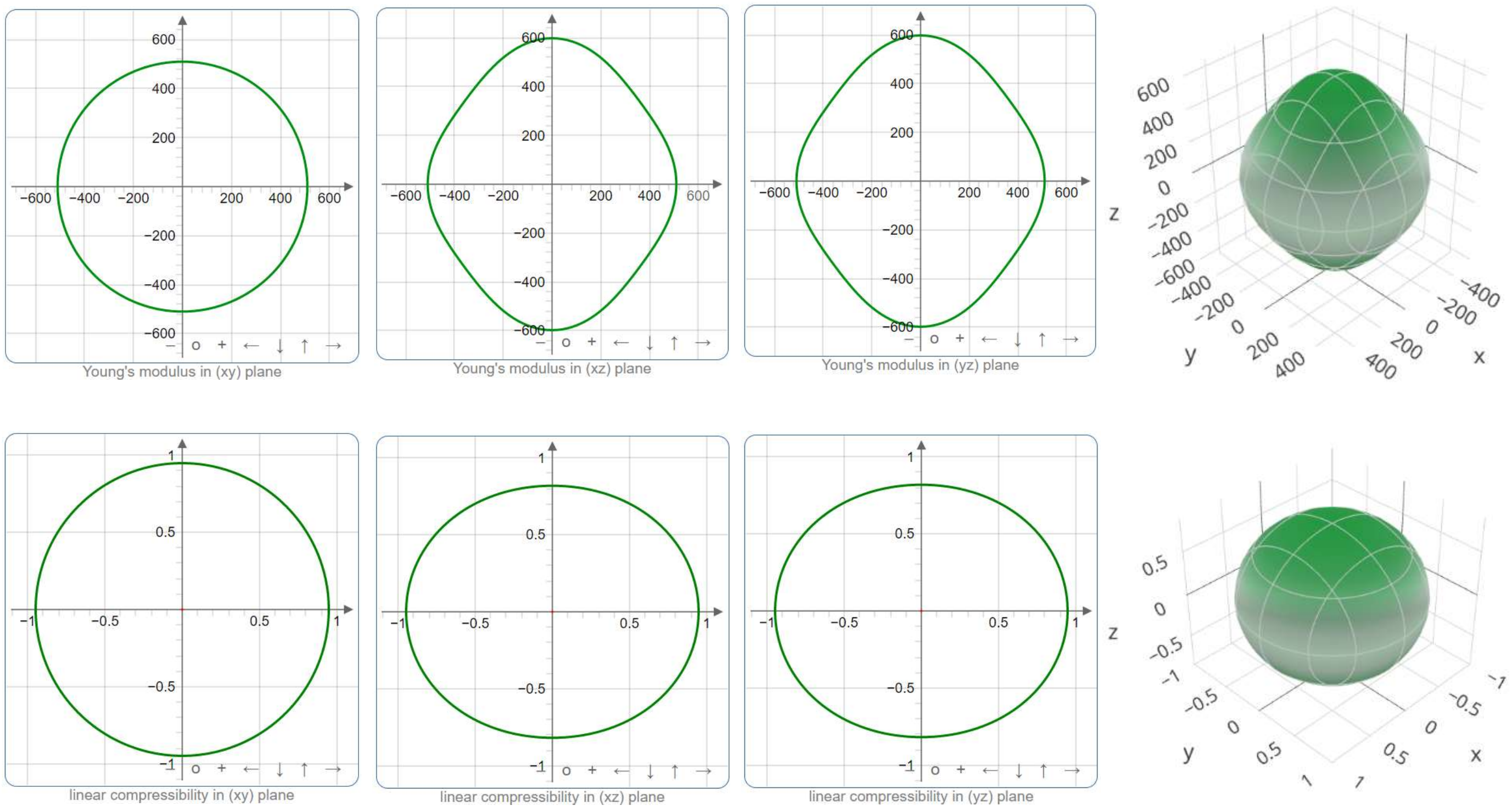

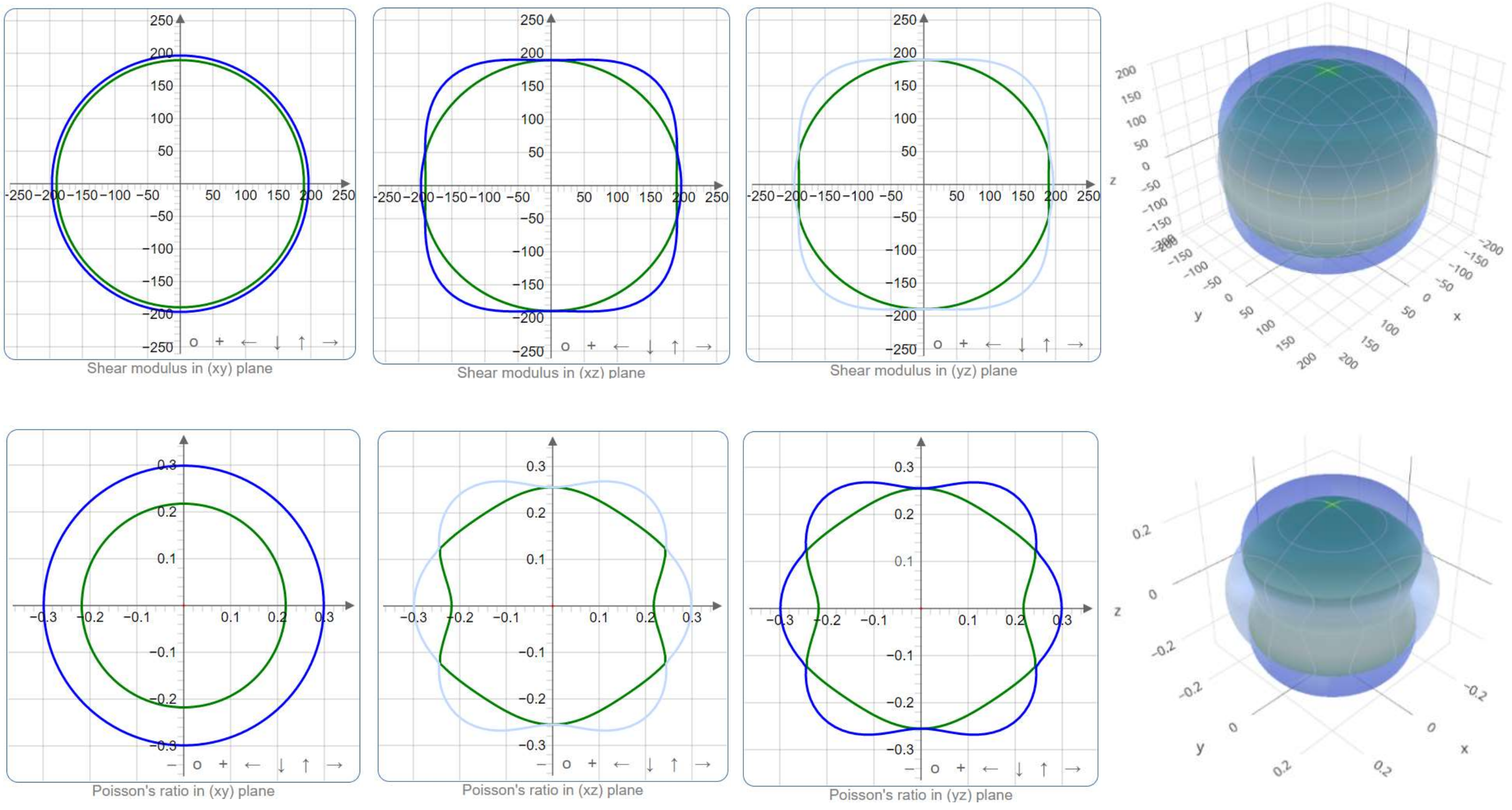

**Figure 3.** Directional dependence of Young's modulus (*E*), linear compressibility (*β*), shear modulus (*G*), and Poisson's ratio (ν) of H-MC.

### 3.4. Acoustic velocities and their anisotropy

The stiffness of a crystal, defined as the resistance to elastic deformation, can be measured by the velocity of sound in it. Sound velocity techniques have increasingly been used to determine the dynamic properties of materials in physics, materials science, seismology, geology, musical instrument designing, and medical science. Moreover, the acoustic velocity in a material is directly related to its thermal conductivity by: $k = \frac{1}{3} C_v l v$. The velocity of the transverse and longitudinal waves through a medium can be estimated from its bulk and shear moduli using the expressions [105]:

$$v_t = \sqrt{\frac{G}{\rho}} \quad \text{and} \quad v_l = \sqrt{\frac{B + 4G/3}{\rho}} \tag{33}$$

where, $\rho$ is the mass-density of the crystal.

The average sound velocity $v_a$ in crystals which can be estimated from the following equation [105]:

$$\nu_a = \left[\frac{1}{3}\left(\frac{2}{\nu_t^3} + \frac{1}{\nu_l^3}\right)\right]^{-\frac{1}{3}} \tag{34}$$

The estimated acoustic velocities of O-MC and H-MC have been enlisted in Table 10.

The acoustic impedance ($Z$) is an important parameter that determines the transfer of sound energy in a medium. The degree of mismatch between the acoustic impedances of the two media explains the amount of reflected and transmitted energy when a sound wave falls at the interface. Sound actually travels more efficiently in materials of high acoustic impedance. Most of the sound gets transmitted or reflected, if the impedance difference of the two materials is about equal or much higher, respectively. Materials with $Z = 0$ and $Z \rightarrow \infty$ are defined as ideal soft and ideal hard surfaces, respectively. Sound energy cannot pass through an ideal soft surface. The acoustic impedance difference between the two media has vast applications in the musical industry, transducer design, acoustic sensor, aerospace industry, industrial factories, automobiles, medical ultrasound imaging, and many underwater acoustic applications. The acoustic impedance of a medium can be evaluated from its shear modulus and density using the following expression [63]:

$$Z = \sqrt{\rho G} \tag{35}$$

Denser and stiffer materials possess higher acoustic impedance. The acoustic impedance of a medium is a function of temperature since density depends on temperature. Calculated values of $Z$ of the compounds are given in Table 10. The noise barriers have become popular in the United States and many European countries over the past few decades. And $Z$ is a significant property for selecting proper materials as noise barriers. Acoustic impudence of materials is needed to evaluate the intensity of the sound reflected at an interface using the reflection coefficient, $R$, as follows [106]:

$$R = \left(\frac{Z_1 - Z_2}{Z_1 + Z_2}\right)^2 \tag{36}$$

Bodies with a pressure reflection factor $R = 1$ will take no energy from the surrounding sound field.

Choosing or designing materials for musical instruments have become a matter of great interest among scientists for a long time. As we know that every vibrating material radiates acoustic energy and the intensity of this radiation is another important parameter for material selection. The frequency of sound (pitch) emitted when an object is struck relates to its material properties and can be measured from $\sqrt{E/\rho}$ [107]. It is crucial to choose proper materials, acoustically bright or dull, depending on the pitch needed. Some materials are suitable for the front plate of a violin, the soundboard of a harpsichord, and the panel of a loudspeaker. The acoustic intensity, $I$, of a material can be defined as [63,108]:

$$I \approx \sqrt{G/\rho^3} \quad (37)$$

where, $\sqrt{G/\rho^3}$ is defined as *radiation factor*. It assesses the ability of materials to efficiently radiate sound. Solids with high *radiation factor* are suitable for soundboards of string musical instruments. One of the most widely used materials for the soundboard of violins is spruce ($I$ = 8.6 $m^4$/kg.s). On the other hand, maple ($I$ = 5.4 $m^4$/kg.s) is used for ribs and back plate of violins whose function is to reflect, not radiate. A high pitch material such as steel is used to make bells. The evaluated radiation factors are listed in Table 10. To give a general idea, relevant acoustic data for a widely used element, silver (Ag) is also given in this table.

The Grüneisen parameter of materials sets the strength of its phonon-phonon interaction. It is in general a function of both volume and temperature. There are different types of Grüneisen constants: acoustic ($\gamma_a$), elastic ($\gamma_e$), lattice ($\gamma_l$), thermodynamic ($\gamma_d$) and electronic ($\gamma_e$). An important fact is that for most metals, as well as for ionic and molecular crystals, the values of $\gamma_a$ and $\gamma_e$ coincide with the value of the thermodynamic Grüneisen constant $\gamma_d$ [110]. On the other hand for some materials, especially rear-earth metals, a considerable difference occurs between $\gamma_a \approx \gamma_e$ and $\gamma_d$. The electronic Grüneisen constant ($\gamma_e$) defines the dimensional changes of a solid in response to the heating of its conduction electrons [111,112]. The elastic Grüneisen constants $\gamma_e$ of the structures under consideration have been calculated from its relation with the Poisson's ratio [113]:

$$\gamma_e = \frac{3(1+\nu)}{2(2-3\nu)} \quad (38)$$

A number of physical processes (thermal conductivity, thermal expansion, temperature dependence of elastic properties, acoustic wave's attenuation) are controlled by this quantity. It also estimates the anharmonic effect of a crystal, e.g., the temperature dependence of phonon frequencies and phonon damping as well as the thermal expansion. The higher is γ, higher is the anharmonicity, the lower is the phonon thermal conductivity. The values of γ obtained for the orthorhombic and hexagonal structures are quite close, 1.66 and 1.60, respectively.

**Table 10.** Density $\rho$ (g/$cm^3$), transverse velocity $v_t$ ($ms^{-1}$), longitudinal velocity $v_l$($ms^{-1}$), average elastic wave velocity $v_a$ ($ms^{-1}$), Grüneisen parameter γ, acoustic impedance $Z$ (Rayl) and radiation factor $\sqrt{G/\rho^3}$ ($m^4$/kg.s) of O-MC and H-MC compounds.

| Compounds | $\rho$ | $v_t$ | $v_l$ | $v_a$ | γ | $Z$ ($\times 10^6$) | $\sqrt{G/\rho^3}$ | Remarks |
|---|---|---|---|---|---|---|---|---|
| O-MC | 9.26 | 4179.44 | 7565.38 | 4617.75 | 1.66 | 38.70 | 0.45 | This work |
| | 9.120 | 3905.00 | 6906.00 | 4343.00 | - | - | - | [17]Expt. |
| | - | 4079.16 | 7522.23 | 4551.58 | - | - | - | [33]Theo. |
| | 9.05 | 4058.9 | 7343.9 | 4522.4 | - | - | - | [34]Theo. |
| H-MC | 11.33 | 4205.97 | 7492.76 | 4641.14 | 1.60 | 47.64 | 0.37 | This work |
| Ag | 10.4 | - | - | - | - | 27.9 | 0.26 | [109]$^{Expt.}$ |

An elastically anisotropic medium suggests anisotropy in sound velocities and vice versa. Therefore, the pure longitudinal and transverse modes are found only along certain directions which are crystal structure dependent. For instance, the pure longitudinal and transverse modes can be found for [100], [010], and [001] directions in an orthorhombic system, whereas, the hexagonal system has only two directions [100] and [001]. The knowledge of these directions and modes is useful for the determination of third-order elastic coefficients from sound velocity measurements in stressed crystalline media [114] and they allow the evaluation of anharmonic properties such as thermal expansion and the interaction of thermal and acoustic phonons. Also all the other modes in any other directions are quasi-longitudinal and quasi-transverse. In the principal directions the acoustic velocities of O-MC and H-MC have been calculated from the single crystal elastic constants as follows [115]:

Orthorhombic:

[100]:

$$[100]\upsilon_l = \sqrt{C_{11}/\rho};[010]\upsilon_{t1} = \sqrt{C_{66}/\rho};[001]\upsilon_{t2} = \sqrt{C_{55}/\rho}$$

[010]:

$$[010]\upsilon_l = \sqrt{C_{22}/\rho}\ ; [100]\upsilon_{t1} = \sqrt{C_{66}/\rho}; [001]\upsilon_{t2} = \sqrt{C_{44}/\rho} \tag{39}$$

[001]:

$$[001]\upsilon_l = \sqrt{(C_{33})/\rho};[100]\upsilon_{t1} = \sqrt{C_{55}/\rho}; [001]\upsilon_{t2} = \sqrt{C_{44}/\rho}$$

Hexagonal:

[100]:

$$[100]\upsilon_l = \sqrt{(C_{11} - C_{12})/2\rho};[010]\upsilon_{t1} = \sqrt{C_{11}/\rho};[001]\upsilon_{t2} = \sqrt{C_{44}/\rho} \tag{40}$$

[001]:

$$[001]\upsilon_l = \sqrt{C_{33}/\rho};[100]\upsilon_{t1} = [010]\upsilon_{t2} = \sqrt{C_{44}/\rho}$$

here, $v_l$ represents the longitudinal sound velocity, $\rho$ is the density. $v_{t_1}$ and $v_{t_2}$ refers the first and second transverse mode, respectively. One may clearly observe that $C_{11}$, $C_{22}$, and $C_{33}$ mainly contributed to the longitudinal waves. The estimated acoustic velocities of the two structures of $Mo_2C$ in principle directions are tabulated in Table 11. Generally, longitudinal velocity is larger than first and second transverse velocities. But for H-MC, the first transverse speed is larger than the longitudinal speed in [100] direction.

**Table 11.** Anisotropic sound velocities ($ms^{-1}$) of O-MC and H-MC along different crystallographic directions.

| Propagation directions | | O-MC | H-MC |
|---|---|---|---|
| [100] | $[100]v_l$ | 7243.53 | 4163.43 |
| | $[010]v_{t1}$ | 4619.82 | 7448.90 |
| | $[001]v_{t2}$ | 4392.55 | 4087.68 |
| [010] | $[010]v_l$ | 7460.08 | - |
| | $[100]v_{t1}$ | 4619.82 | - |
| | $[001]v_{t2}$ | 4094.99 | - |
| [001] | $[001]v_l$ | 7482.11 | 7922.61 |
| | $[100]v_{t1}$ | 4392.55 | 4087.68 |
| | $[001]v_{t2}$ | 4094.99 | 4087.68 |

The anisotropy in sound velocity reflects the anisotropy in atomic bondings in O-MC and H-MC structures.

## 3.5. Electronic properties

### *3.5.1. Electronic band structure*

The electronic band structure is an important feature that helps to explain the electronic, optical, and magnetic properties of materials at the microscopic level. It also helps to calculate the effective masses of charge carriers. It also determines the bonding characteristics to a large extent. However, a material's charge transport properties can be understood in greater details if one can identify the character of dominant bands near the Fermi level. The electronic energy band structures of the orthorhombic and hexagonal structures of $Mo_2C$, calculated with energy smearing widths of 0.2 eV and 0.1 eV, along the high symmetry directions of the first Brillouin zone are shown in Figure 4(a) and 4(b), respectively. The horizontal dotted line represents the Fermi level ($E_F$). It is observed that the bands for O-MC are more localized than those for H-MC. Due to the overlap of valence band and conduction bands (see Figs. 3 and 4) there is no band gap. Thus, both the structures exhibit metallic character. The bands crossing the Fermi level are shown in different colors with their corresponding band numbers. Both electron-like and hole-like features along different directions in the BZ can be observed for each compound. The band structure calculation also helps us to grasp the shape of the underlying Fermi surfaces.

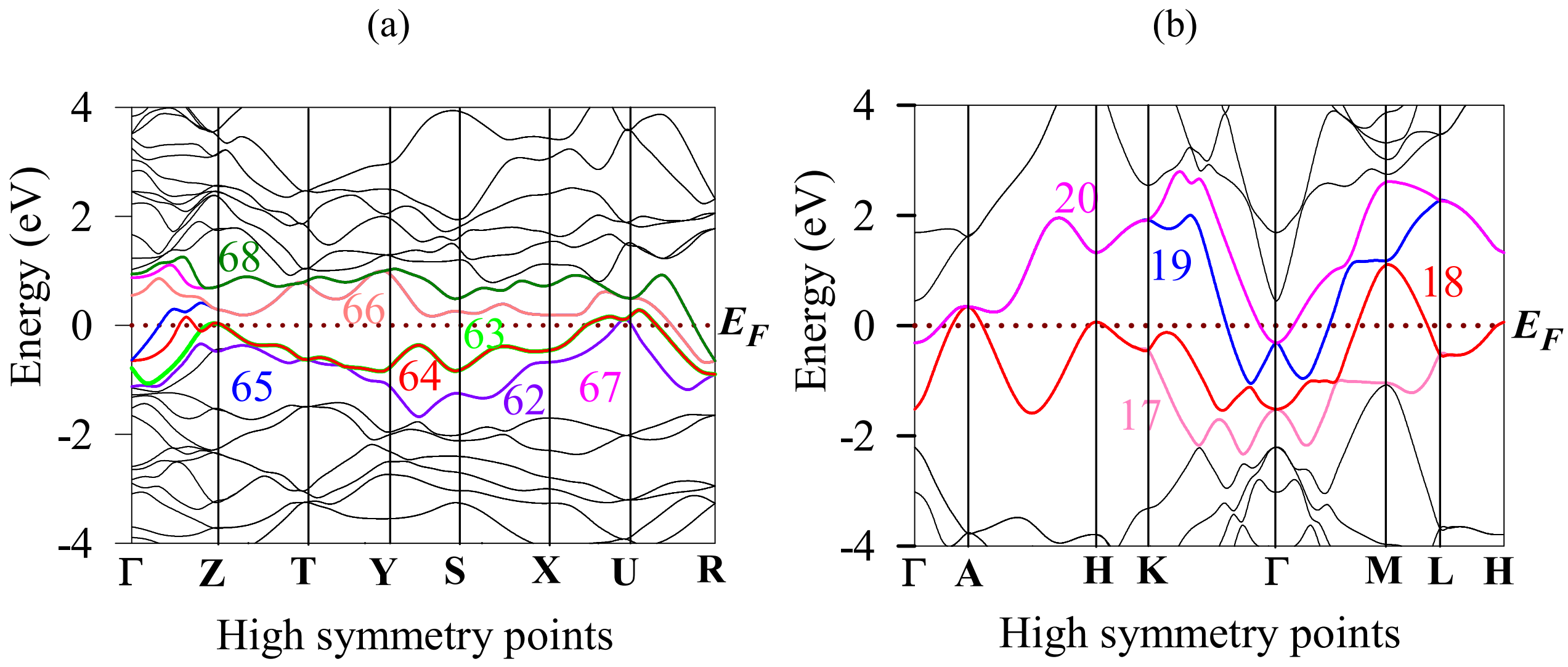

**Figure 4.** Electronic band structure of (a) O-MC and (b) H-MC along several high symmetry directions of the Brillouin zone in the ground state.

### *3.5.2. Density of states*

The bonding nature between atoms and the contribution of different atoms/orbitals to the optoelectronic properties of a material can be better understood by studying its total and partial density of states (DOS). Besides, the electronic heat capacity and spin paramagnetic susceptibility are two important physical quantities directly related to the electronic density of states at the Fermi level, N($E_F$) [116]. Figure 5(a) and 5(b) show the calculated total and partial DOS of O-MC and H-MC, respectively. The Fermi level is represented by the vertical broken line at zero energy. The finite values of TDOS at the Fermi level for both compounds disclose their metallic nature. Due to significantly higher TDOS at Fermi level, the orthorhombic crystal predicts better electrical conductivity. The calculated total density of states (TDOS) at Fermi energy of O-MC and H-MC at $E_F$ are 6.28 and 1.49 states per eV per unit cell, respectively. PDOS calculations show that near the Fermi level the main contribution comes from Mo 4*d* states for both crystal symmetries. There is significant hybridization between Mo 4*d* and C 2*p* electronic states for both structures. Such hybridization near the Fermi energy is often indicative of formation of strong covalent bonding.

The electron-electron interaction parameter (the so-called Coulomb pseudopotential) is estimated using the following relation [117]:

$$\mu^* = \frac{0.26N(E_F)}{1 + N(E_F)} \tag{41}$$

The calculated Coulomb pseudopotential of O-MC and H-MC are 0.22 and 0.15, respectively. The transition temperature, $T_c$ of superconducting systems is adversely affected by the repulsive Coulomb pseudopotential [117-119] which inhibits electronic pairing.

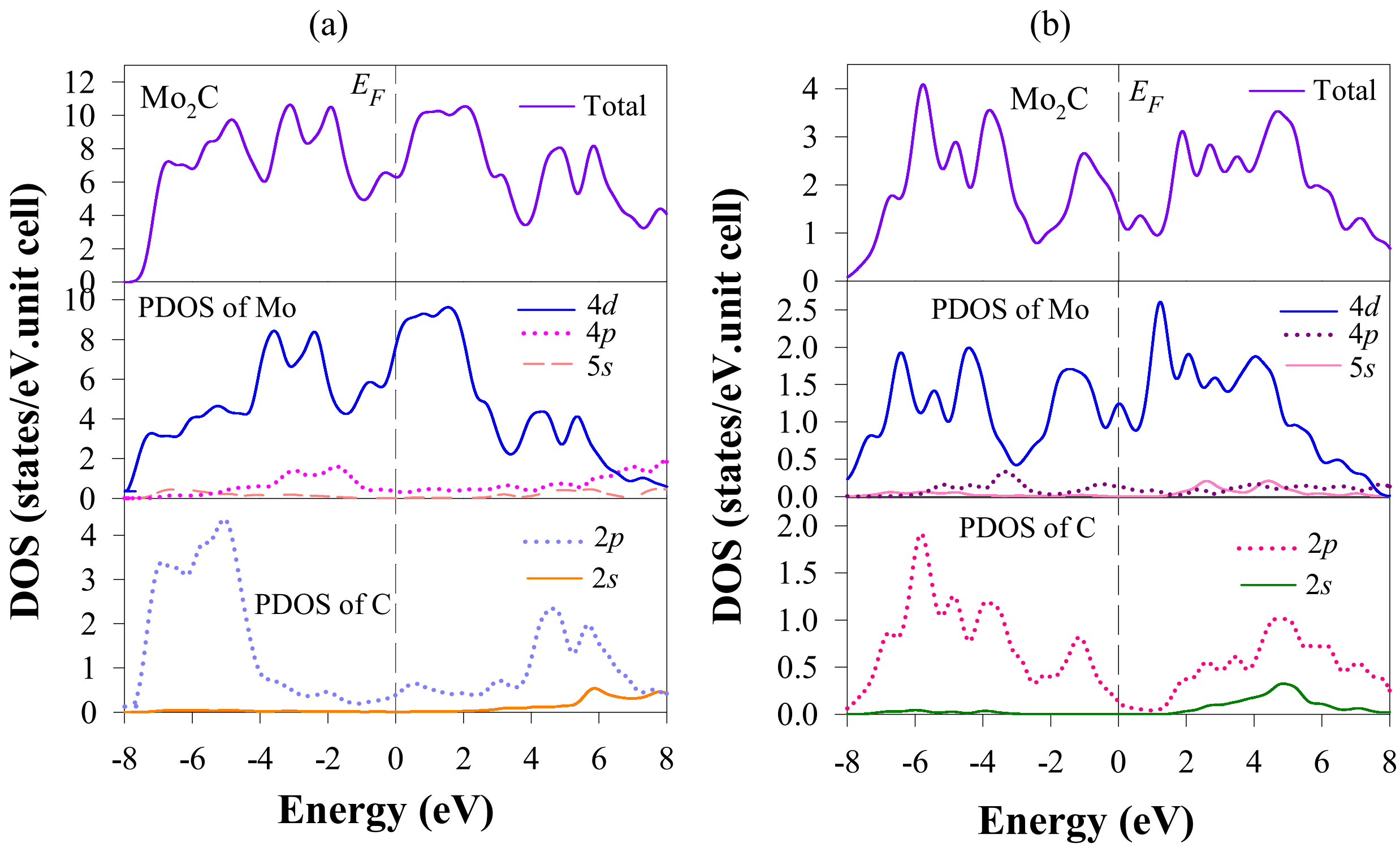

**Figure 5.** Total and partial electronic density of states (TDOS and PDOS) of (a) O-MC and (b) H-MC as a function of energy. The Fermi level is placed at zero energy.

### *3.5.3. Charge density distribution*

The electronic charge distribution map is a useful tool to assess the nature of interatomic chemical bonds. It shows accumulation and depletion of electronic charges close to different atomic species. The accumulation of charges between two atoms illustrates covalent bonding between them. The existence of ionic bonds can be predicted from a negative and positive charge balance at the atom positions. Thus, to get a better understanding of chemical bonding between the ions of O-MC and H-MC compounds, we have studied the electronic charge density distribution shown in Figures 6 and 7, respectively. The adjacent color scales illustrate the intensity of total electronic density in the unit of e/Å$^3$. The red and blue colors represent high and low electronic densities, respectively. The charge density value ranges from 0.1437 to 12.71 (electrons) for orthorhombic and 0.1724 to 12.74 (electrons) for hexagonal structures. The charge density distribution maps of O-MC, 3D view in the (101) plane, show clear signature of covalent bonding between C-Mo and Mo-Mo atoms. In some of the planes (not shown here), Mo atoms have high electron density compared to C atoms and ionic bondings are expected. Figure 7 shows similar features. These findings for both structures are in accord with the Mulliken bond population analysis. Both O-MC and H-MC compounds show direction and plane dependency of charge density distribution.

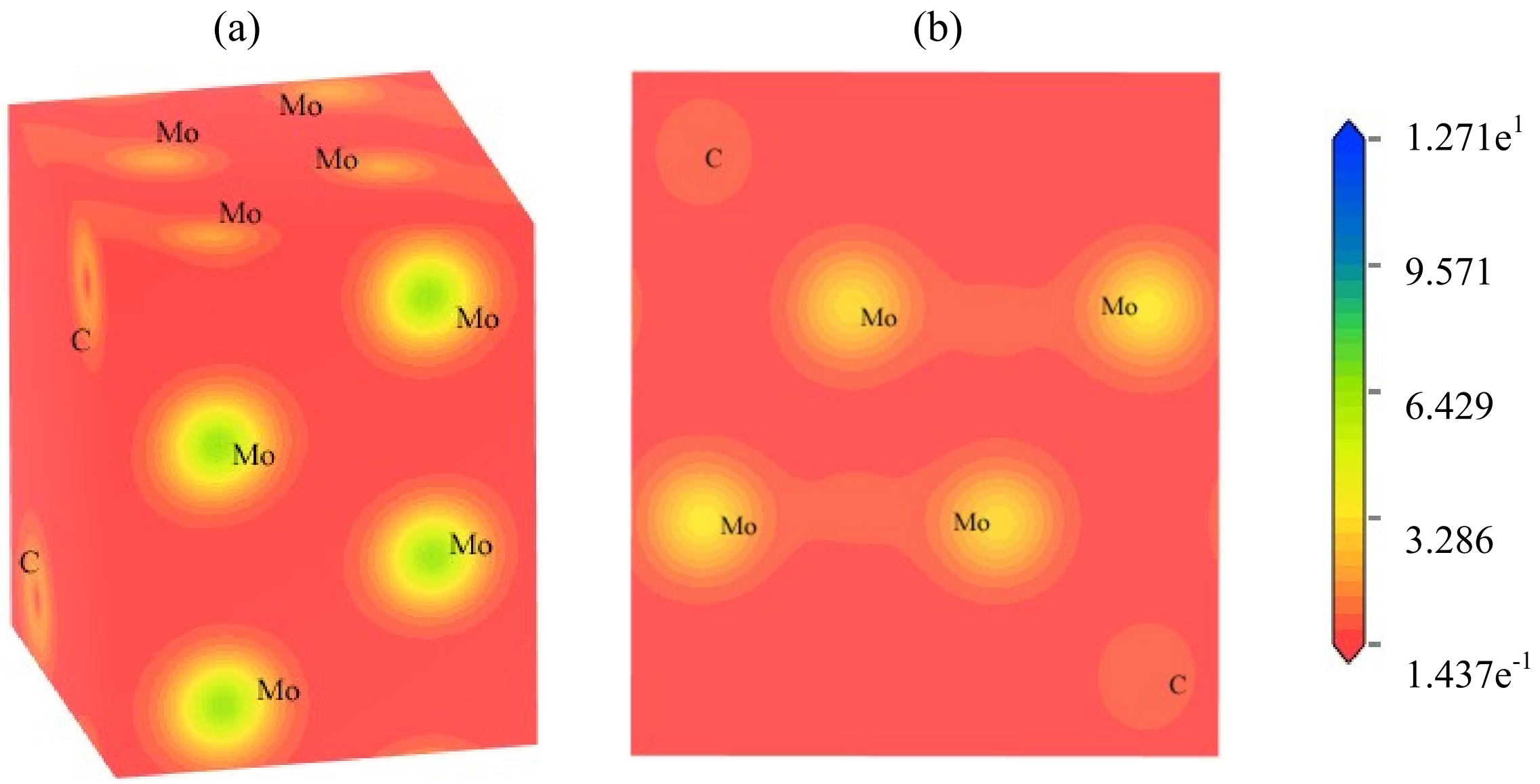

**Figure 6.** Charge density distribution in various crystal planes. (a) 3D view and (b) in the (101) plane of O-MC. The charge density scale is shown on the right.

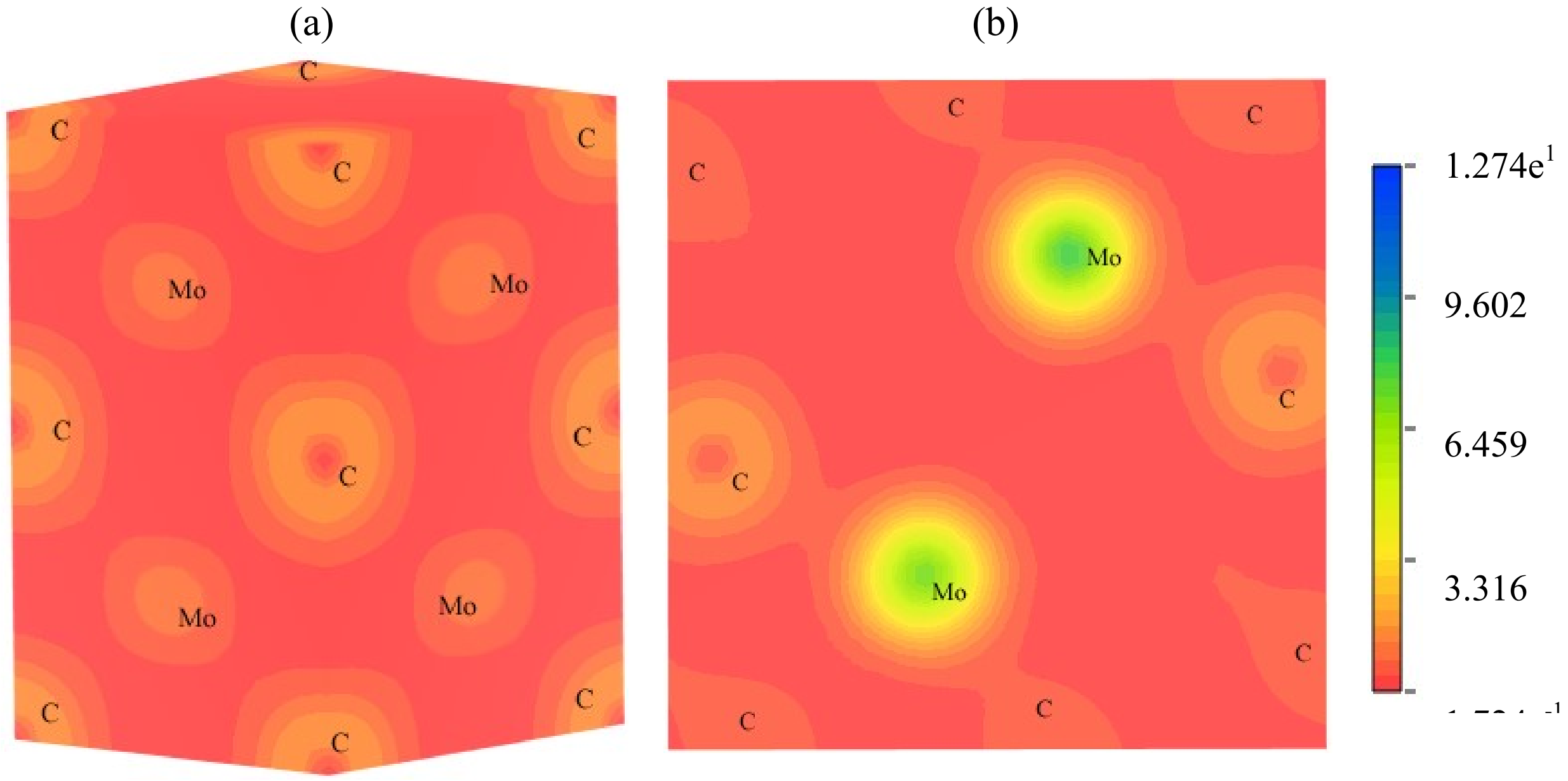

**Figure 7.** Charge density distribution in various crystal planes. (a) 3D view and (b) in the (111) plane of H-MC. The charge density scale is shown on the right.

### *3.5.4. Fermi surfaces*

The study of Fermi surface is crucial for understanding the behavior of occupied and unoccupied electronic states of a metallic material at low temperatures. Several properties, such as electronic, optical, thermal, and magnetic, strongly depend on Fermi surface topology. The electrons near the Fermi surface are involved in formation of superconducting state. Therefore, it is important to figure out its nature. The 3D plot of the Fermi surfaces of O-MC and H-MC for the bands crossing Fermi levels are shown in Figures 8 and 9, respectively. A tiny electron-like sheet appears around *U*-point for bands 62 and 63 of O-MC. For band 64, an electron-like sheet is found at around *Z*-point. Hole-like sheets appear around the *Γ*- and *R*-points for band 65. The Fermi surfaces for bands 66, 67, and 68 of O-MC are quite similar. Therefore, O-MC contains both electron- and hole-like sheets. These findings are in complete agreement with previous study [33]. On the other hand, there are 4 bands (17, 18, 19, and 20) crossing the Fermi level in H-MC. For band 17, there are tiny electron-like sheets around the *A*- and *H*-points. For band 18, electron-like sheets appear around *A*- and *M*-points. Electron-like sheets appear along the *Γ*-*A* path for band 19 and 20. So, H-MC structure contains only electron-like sheets.

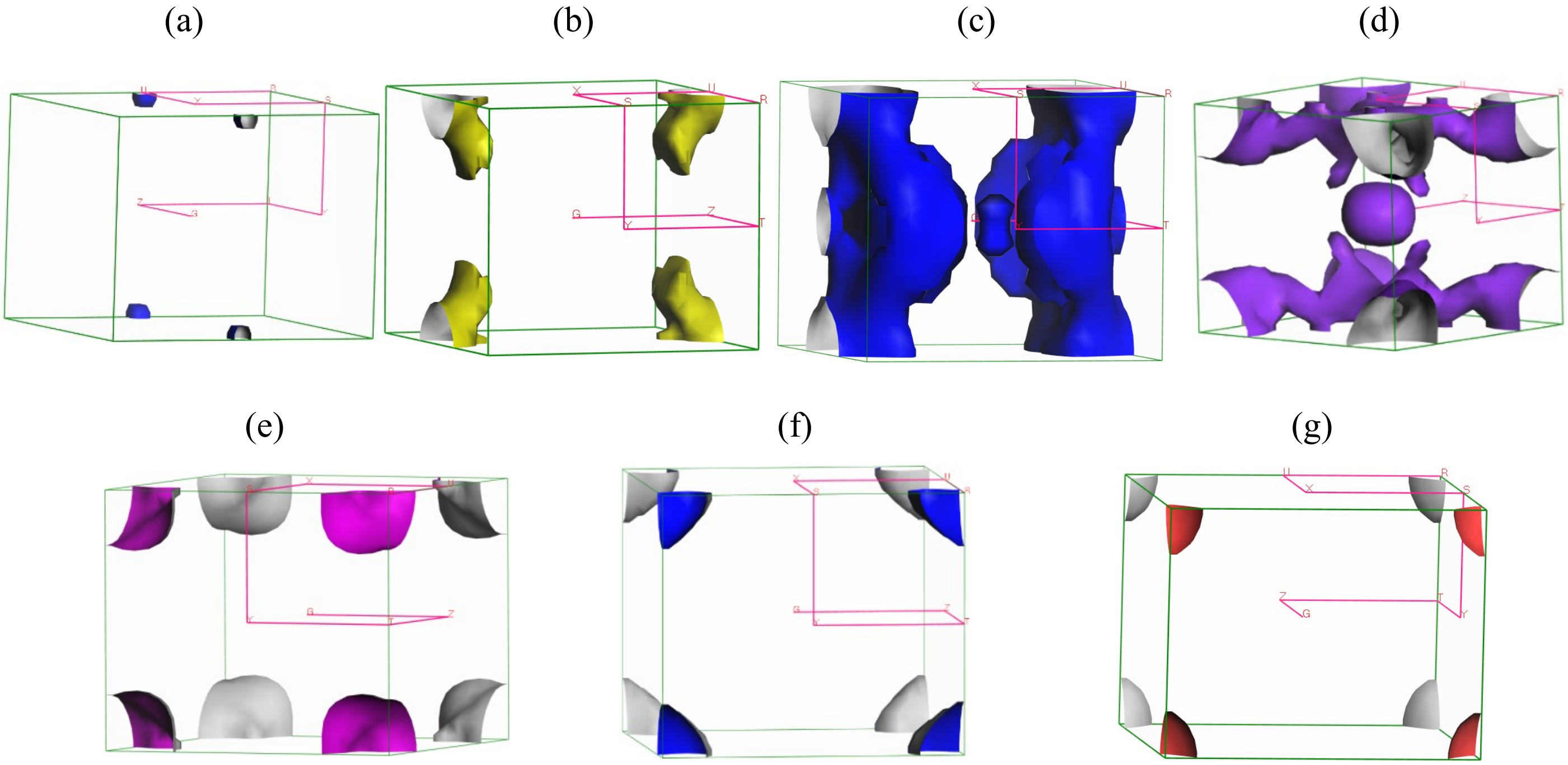

**Figure 8.** Fermi surface of O-MC for bands (a) 62, (b) 63, (c) 64, (d) 65, (e) 66, (f) 67, and (g) 68, respectively.

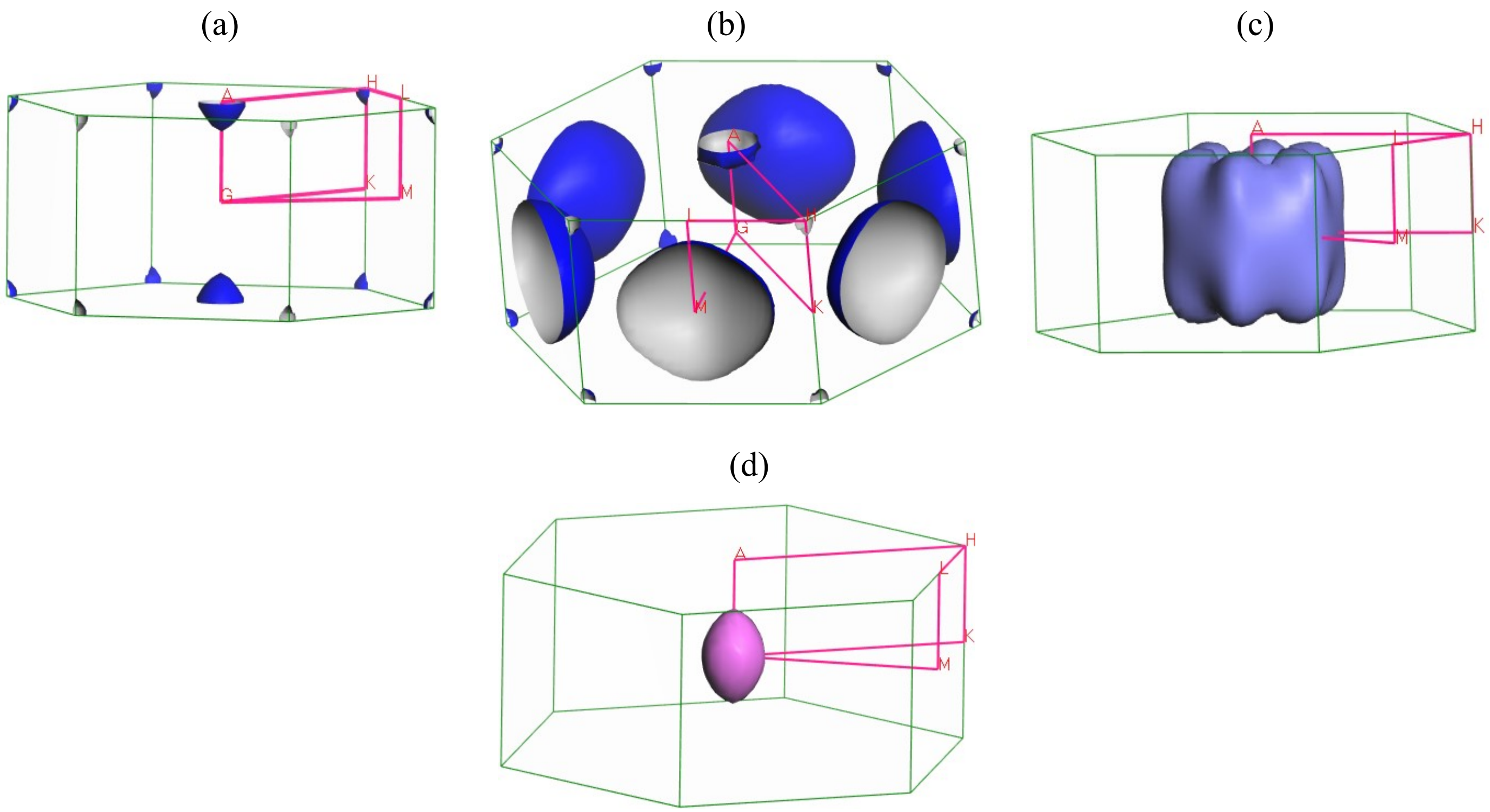

**Figure 9.** Fermi surface of H-MC for bands (a) 17, (b) 18, (c) 19 and (d) 20, respectively.

### 3.6. Vibrational properties

Many dynamical properties of solids are affected directly or indirectly by the phonon dispersion spectra and phonon density of states (PDOS) [120]. The phonon dispersion spectra (PDS) has become an important concept to get many information regarding dynamic lattice stability/instability, phase transition and vibrational contribution to heat conduction, thermal expansion, Helmholtz free energy, and heat capacity in solids [28]. The electron–phonon interaction function is also directly related to the PDOS. Many of the optical properties of materials are controlled by their optical phonons. The density functional perturbation theory (DFPT) finite displacement method (FDM) [121,122] has been employed to calculate phonon dispersion spectra (PDS) and total phonon density of states (PDOS) of $Mo_2C$. The phonon dispersion and density of states of O-MC and H-MC phases are shown in Figures 10(a) and 10(b), respectively. The dispersion curves are shown along the high-symmetry directions of the Brillouin zone. The dynamical stabilities of O-MC and H-MC structures are investigated from the phonon dispersion spectra. A compound is dynamically stable if the phonon frequencies over the whole BZ are positive. The presence of imaginary (negative) phonon frequency ensures the presence of soft phonon modes and dynamic instability. Both the phases are free of imaginary vibrational frequencies in the whole BZ, which is a sign of their dynamical stability. The total number of phonon modes is three times the total number of atoms per unit cell. A unit cell consist of *N* atoms possesses 3 acoustic mode and (3*N*-3) optical modes. The acoustic modes contain one longitudinal and two transverse acoustic branches. Acoustic phonon is caused by the coherent vibrations of atoms in a lattice outside their equilibrium position. Conversely, the optical phonon is originated due to the out-of-phase oscillations of the atoms in the lattice when an atom moves to the left and its neighbor to the right. Acoustic phonons contribute to sound propagation in crystals and are related to the stiffness of crystals. The unit cell of orthorhombic structure consists of 12 atoms, creating a total of 36 normal lattice vibration modes, including 3 acoustic (pink) and 33 optical modes. Optical phonons are phonons with non-zero frequencies at the $\Gamma$-point. Hexagonal unit cell contains 4 atoms in total which produce 12 vibrational modes with 3 acoustic (pink) and 9 optical branches. The rapid flattening of the dispersion of the acoustic modes away from the center of the BZ is observed for orthorhombic structure. The optical behavior of a material strongly depends on the optical branches. Sometimes the optical modes are separated into two branches: lower and upper optical branches. There is a frequency gap between two branches for both phases due to the atomic mass difference between Mo and C atoms. This gap is broad, and the branches are more localized for orthorhombic structure. In both cases, the lower optical branches overlap with the acoustic branches indicating the absence of phononic bandgap between the acoustic and optical branches. It also predicts the suitability of the thermal transport of the studied compounds. For O-MC and H-MC, the highest phonon frequency for optical mode appears around $\Gamma$ point, and the values are 21.5 THz and 23.05 THz, respectively.

We have also calculated the total and atomic partial PHDOS for each compound, displayed alongside the PDCs, to see the contribution of each band to different atomic modes of vibrations. The PDOS revealing that the acoustic and the lower optical modes arise due to the vibration of heavier Mo atoms for both O- and H-MC. The higher optical branches (with frequencies > 18.6 THz and > 15.7 THz for Ortho- and Hexa-phases, respectively) mainly originate from the vibration of lighter C-atoms. In both cases, the PDOS curves display gaps between the two branches. The prominent peaks in the PHDOS are observed around 18.8 and 17.6 THz for O-MC and H-MC, respectively. The flatness of the bands produces the peaks in PHDOS and larger dispersion of bands (both upward and downward going bands) decrease the heights of peaks in the total PHDOS.

According to the group theory [123], there are a total of 36 and 8 vibrational modes in O-MC and H- MC, respectively. The vibrational modes at the $\Gamma$-points are assigned as follows:

$$\Gamma = 5B_{1u} + 4B_{2u} + 5B_{3u} + 4A_g + 4A_u + 5B_{1g} + 4B_{2g} + 5B_{3g} \quad (42)$$

and

$$\Gamma = 2A_{2u} + B_{1u} + B_{2g} + E_{2g} + 2E_{1u} + E_{2u} \quad (43)$$

For orthorhombic structure, among all these vibrational modes the first three modes ($B_{1u}$, $B_{2u}$ and $B_{3u}$) are acoustic with zero frequencies at the $\Gamma$-point. The other modes include 18 Raman- and 11 infrared (IR)-active optical modes. Rests of the modes are optically inactive. For hexagonal structure, $A_{2u}$ and $2E_{1u}$ are three acoustic modes with zero phonon frequency at the $\Gamma$-point. The optical modes possess 2 Raman- and 3 IR-active optical modes. The remaining modes ($B_{2g}$, $2E_{2u}$, and $B_{1u}$) are optically inactive.

Since both the structures contain a center of symmetry, the rule of mutual exclusion applies; no Raman active bands should be infrared-active, and vice versa. The 18 and 2 expected Raman active phonon modes for O-MC and H-MC, respectively, are shown below:

$$\Gamma_{Raman} = 4A_g + 5B_{1g} + 4B_{2g} + 5B_{3g} \quad (44)$$

and

$$\Gamma_{Raman} = 2E_{2g} \quad (45)$$

Table 11 lists the calculated frequencies and symmetries of all active Raman modes of O-MC and H-MC with their available experimental and calculated values. Our calculated values are in good agreement with the available results. Table 11 and Equation 44 show that the most observed symmetry are $B_{1g}$ and $B_{3g}$.

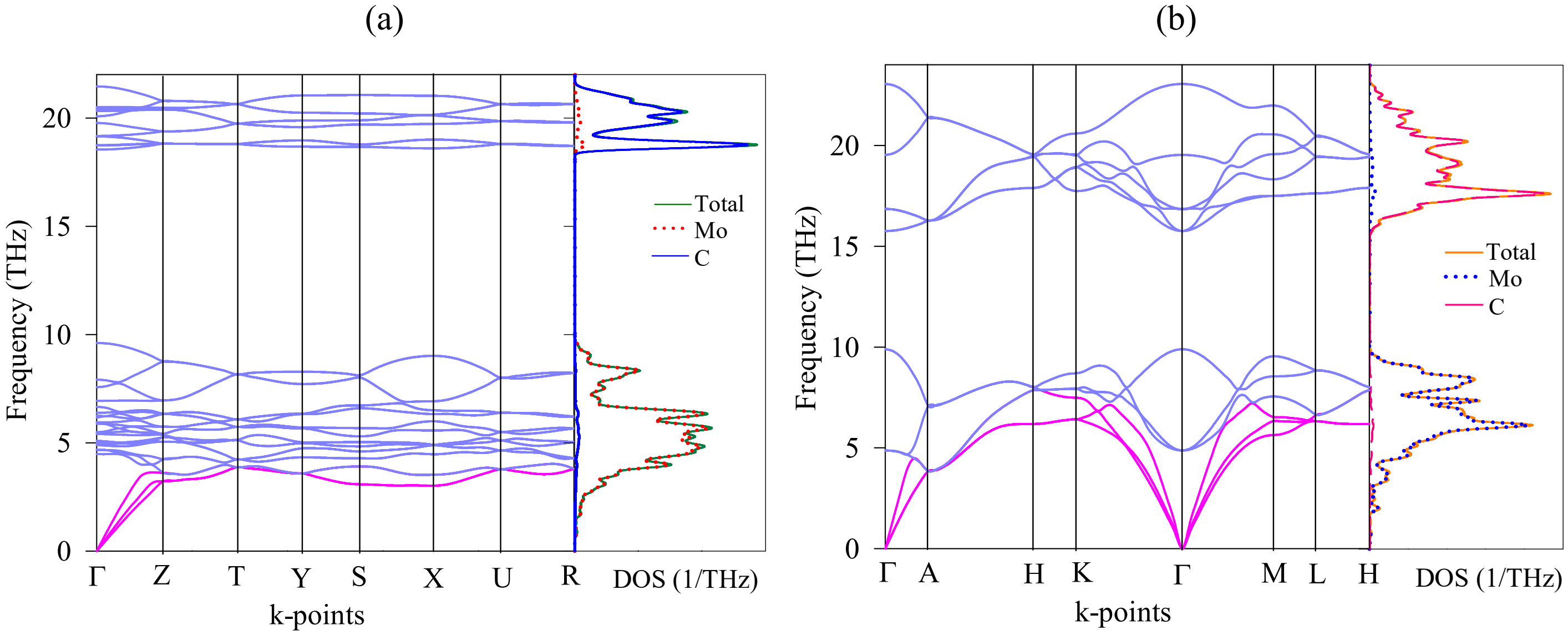

**Fig 10.** Calculated phonon dispersion spectra and phonon DOS of (a) O-MC and (b) H-MC in the ground state.

**Table 11.** Calculated values of Raman-active optical modes at $\Gamma$ point for orthorhombic and hexagonal phase of $Mo_2C$.

| Compounds | Theoretical mode symmetry | Raman shift ($cm^{-1}$) | | |
|---|---|---|---|---|
| | | This work | [37]Theo. | [37]Expt. |
| O-MC | $B_{3g}$ | 149.541 | 147.4 | 143.4 |
| | $B_{1g}$ | 155.945 | 151.6 | - |
| | $A_g$ | 156.480 | 152.5 | 143.4 |
| | $B_{3g}$ | 167.139 | 167.6 | 170.6 |
| | $A_g$ | 181.858 | 192.1 | 182.5 |
| | $B_{2g}$ | 183.176 | 185.8 | - |
| | $B_{2g}$ | 198.530 | 198.4 | - |
| | $B_{1g}$ | 198.532 | 195.6 | - |
| | $B_{3g}$ | 213.111 | 215.3 | 213.0 |
| | $B_{1g}$ | 222.336 | 221.9 | - |
| | $A_g$ | 231.523 | 238.8 | 235.8 |
| | $B_{2g}$ | 320.702 | 315.2 | - |
| | $B_{1g}$ | 618.498 | 605.0 | - |
| | $B_{3g}$ | 625.360 | 607.1 | - |
| | $A_g$ | 659.610 | 647.4 | 652.6 |
| | $B_{2g}$ | 682.750 | 681.2 | - |
| | $B_{1g}$ | 684.156 | 673.2 | - |
| | $B_{3g}$ | 715.756 | 707.6 | - |
| H-MC | $E_{2g}$ | 162.254 | - | - |
| | $E_{2g}$ | 162.254 | - | - |

**3.7. Optoelectronic properties**

The optical study of solids has been drawing great interest in current science and technology. Optical materials have widespread applications as display devices, lasers, photodetectors, solar cells, sensors, reconfigurable photonics, etc [124]. Interaction of material with incident electromagnetic radiation is controlled by its optical properties. For optoelectronic devices, the main focus is the response of the devices to visible light. The optical properties of solids can be completely studied from various energy (frequency) dependent properties, namely, dielectric function, refractive index, loss function, conductivity, reflectivity, and absorption coefficient. Some popular optical devices, such as LCD screens, 3D movie screens, polarizers, wave plates, are based on optical anisotropy. Both O-MC and H-MC show mechanical anisotropy, therefore we have calculated optical parameters of both structures for photon energies up to 30 eV for two polarization directions [100] and [001] of the incident electric field. The optical studies are carried out for the first time. Both electronic band structures and density of states calculation show the metallic nature of O-MC and H-MC. The dielectric function (real and imaginary parts) enlightens various optical properties of materials completely, such as lattice vibrations, free carrier absorption, superconducting gaps, plasmon resonances, chemical bonding, excitons, or inter-band absorption, intra-band transition [125–127].The Drude damping correction is required for metallic materials [128,129]. The calculations of the optical constants of our compounds have been done using screened plasma energy of 10 eV and a Drude damping of 0.2 eV as prescribed in the CASTEP. The frequency dependent real $\varepsilon_1(\omega)$ and imaginary $\varepsilon_2(\omega)$ parts of the dielectric function in the energy range 0-30 eV have been calculated for both compounds (Figure 11). It is well known that the real part of this function $\varepsilon_1(\omega)$ explains the speed of light in the material and polarizability, whereas the imaginary part, $\varepsilon_2(\omega)$, is related to the band structure and absorptive behavior of materials. In metals, intraband transitions dominate at low energy and the interband term strongly depends on the electronic band structure [100]. The static dielectric constant $\varepsilon_1(0)$ is defined as the real part of dielectric constant $\varepsilon_1$ at zero frequency. Figure 11 shows that the $\varepsilon_2(\omega)$ approaches zero at around 28.25 and 28.87 eV in the ultraviolet energy region for O-MC and H-MC, respectively. Photon absorption occurs in the energy region where $\varepsilon_2(\omega)$ is nonzero. Therefore, both compounds are transparent above ~29 eV. The imaginary parts of O-MC in Figure 11(a) show isotropy. The large anisotropy behavior of both dielectric functions is observed for H-MC.

The index of refraction of a material is an important parameter measuring its suitability as optical devices, such as photonic crystals and waveguides. The refractive indices of studied structures are estimated from following equation: $N(\omega) = n(\omega) + ik(\omega)$, where $k(\omega)$ is the extinction coefficient. The refractive index ($n$) and the extinction coefficient ($k$) describe the phase velocity and absorption of light in the medium, respectively. The extinction coefficient has great interest for optoelectronic devices. Figure 12 illustrates the refractive indices as a function of photon energy in two polarization directions for both O-MC and H-MC. It gradually decreases in the high energy region for both structures. H-MC shows more anisotropy than O-MC. For both O-MC and H-MC, the refractive indices at low photon energies are high, irrespective of the state of polarization of the incident light.

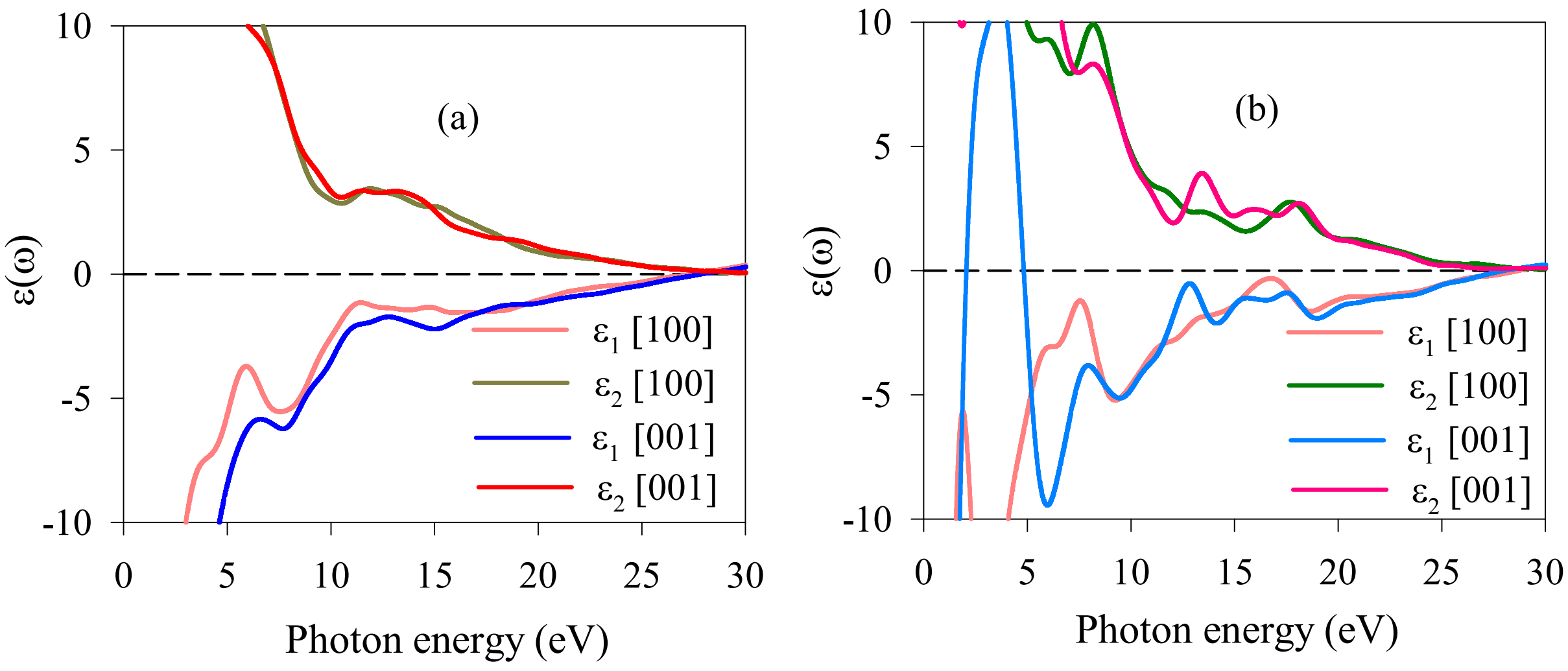

**Figure 11.** Photon energy dependence of real part, $\varepsilon_1(\omega)$ and imaginary part, $\varepsilon_2(\omega)$ of the dielectric functions for (a) O-MC and (b) H-MC.

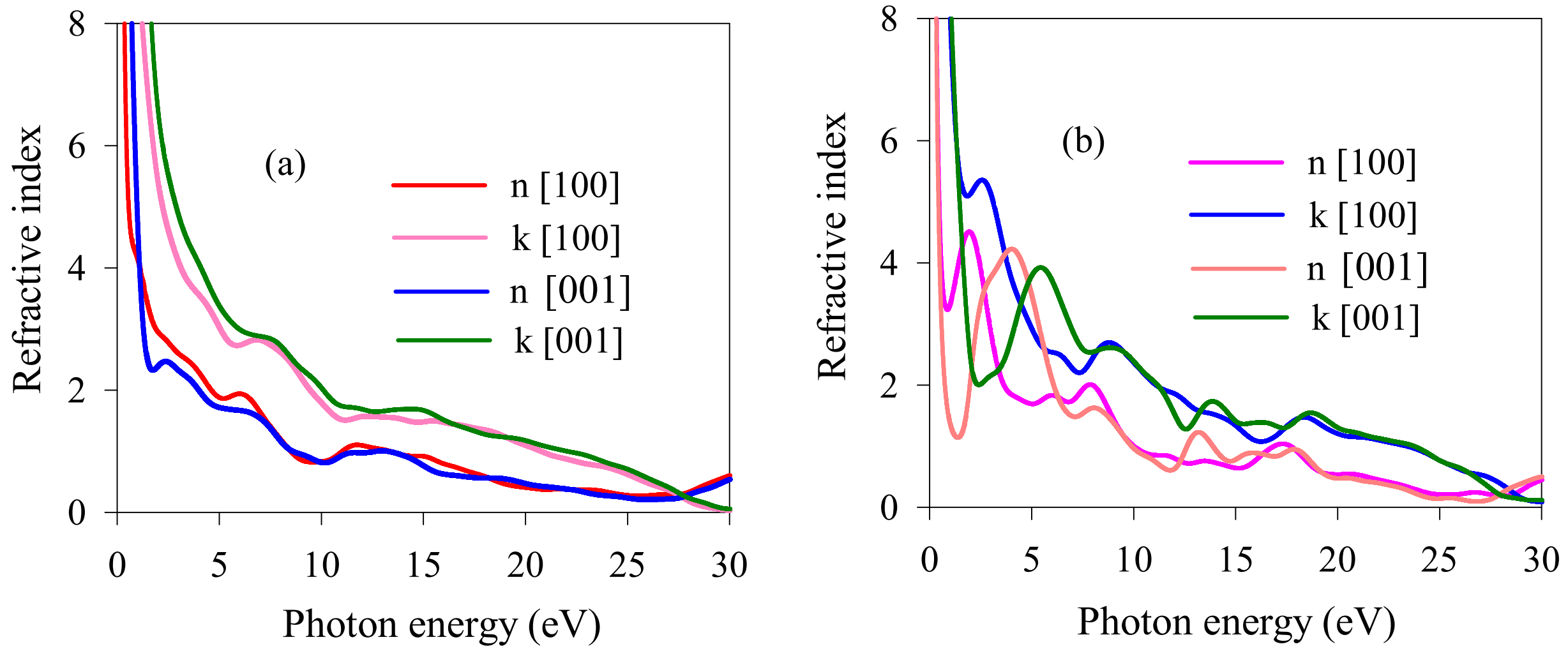

**Figure 12.** Photon energy dependence of refractive index for (a) O-MC and (b) H-MC.

A matter absorbs incident photon energy via its constituent atoms and electrons. The ability of a material to absorb incident electromagnetic wave can be inferred from the absorption coefficient ($\alpha$). The photon absorption in materials increases for the high absorption coefficient and thereby exciting the electrons from valence band to conduction band. It also characterizes material's nature, such as metallic, semiconducting, or insulating. Figure 13 illustrates the energy-dependent absorption spectra along two polarization directions [100] and [001] for both O-MC and H-MC. It is noticed that both structures show absorption anisotropy. Absorption starts at zero photon energy for both the polymorphs in both polarization directions. Therefore, both compounds are metal, which is consistent with the band structure and DOS studies. The maximum absorption for O-MC appears around 15.5 eV and 18.8 eV for [100] and [001] polarizations, respectively. The maximum absorption for H-MC appears around 18.4 eV and 19.1 eV along [100] and [001] polarizations, respectively. The anisotropy is drastically reduced at energies around 23.6 eV and 24.2 eV for O-MC and H-MC, respectively.

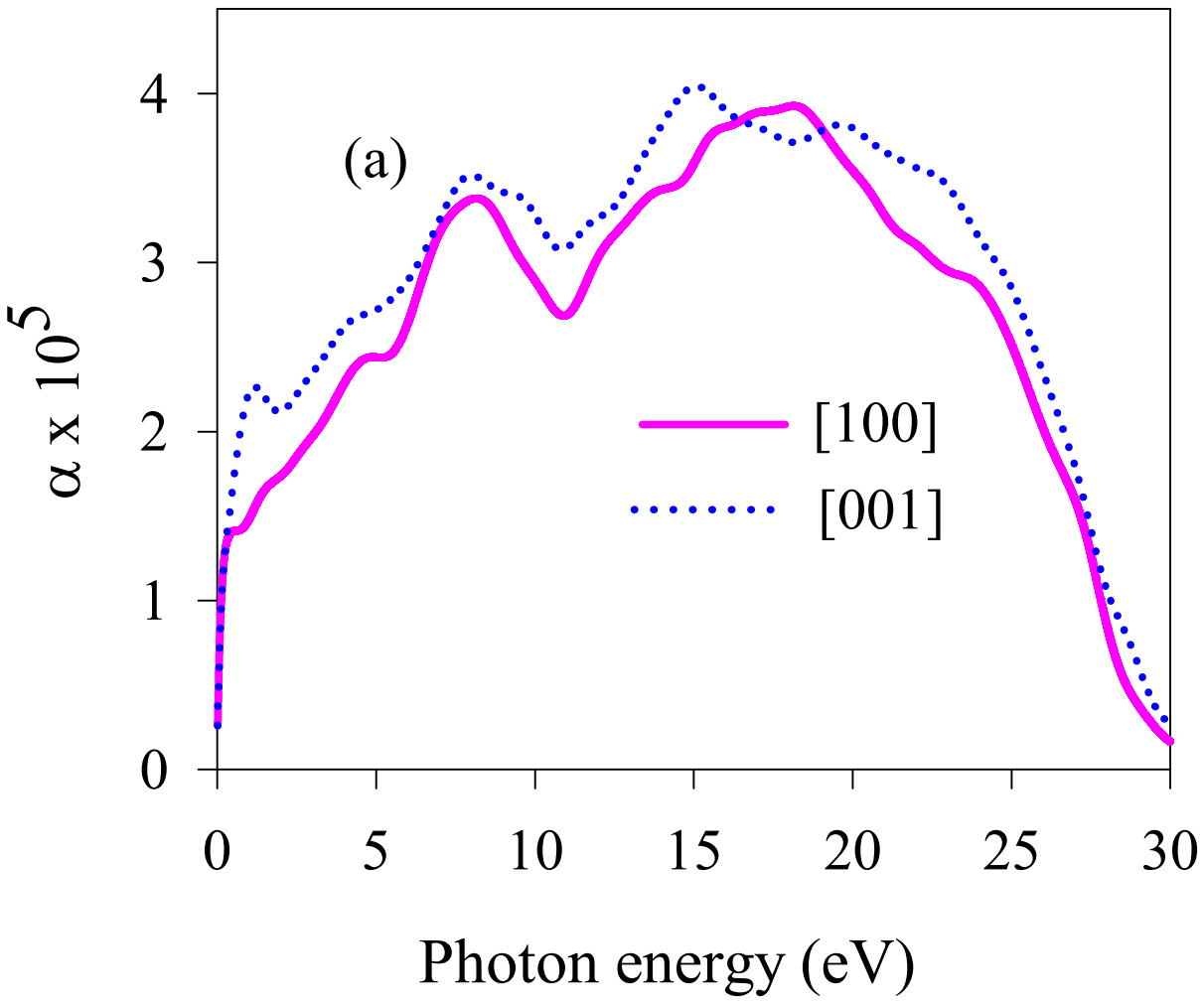

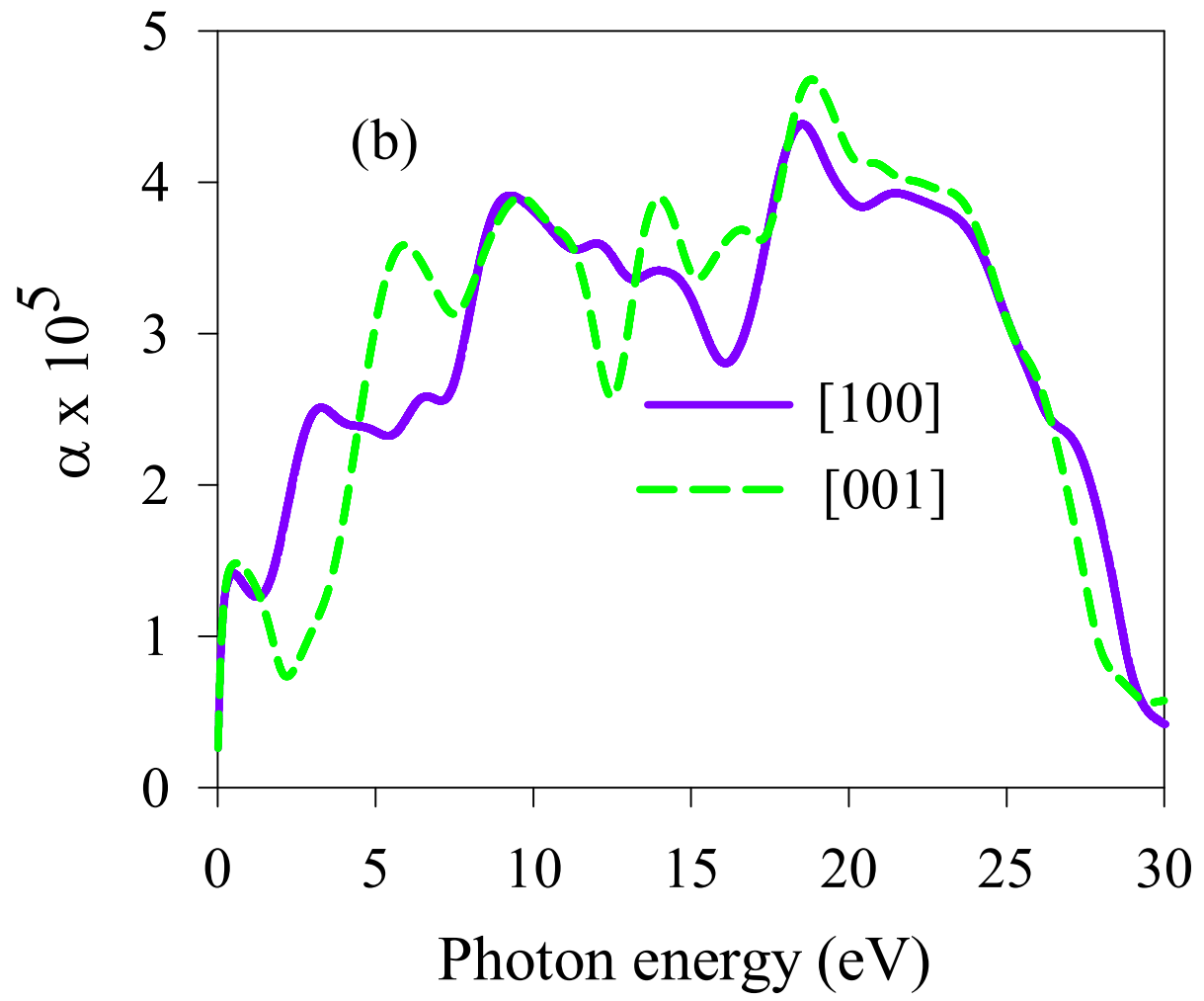

**Figure 13.** Photon energy dependence of absorption coefficient, $\alpha(\omega)$ for (a) O-MC and (b) H-MC.

Figures 14(a) and 14(b) illustrate the real part of the photoconductivity (σ) spectra of O-MC and H-MC, respectively. The nonzero photoconductivity at zero photon energy for both compounds manifests the absence of electronic band gap, which is obvious from electronic band structure calculations. The maximum σ for both compounds along both directions of polarization is at zero photon energy. O-MC shows strong anisotropy in optical conductivity. On the other hand, H-MC shows strong anisotropy in low energy regions and isotropy above 17 eV. Generally, the intraband contribution to the optical properties dominates in the low energy (infrared) part of the spectra. On the other hand, peaks in absorption and conductivity spectra in the high energy part arise due to the interband transition.

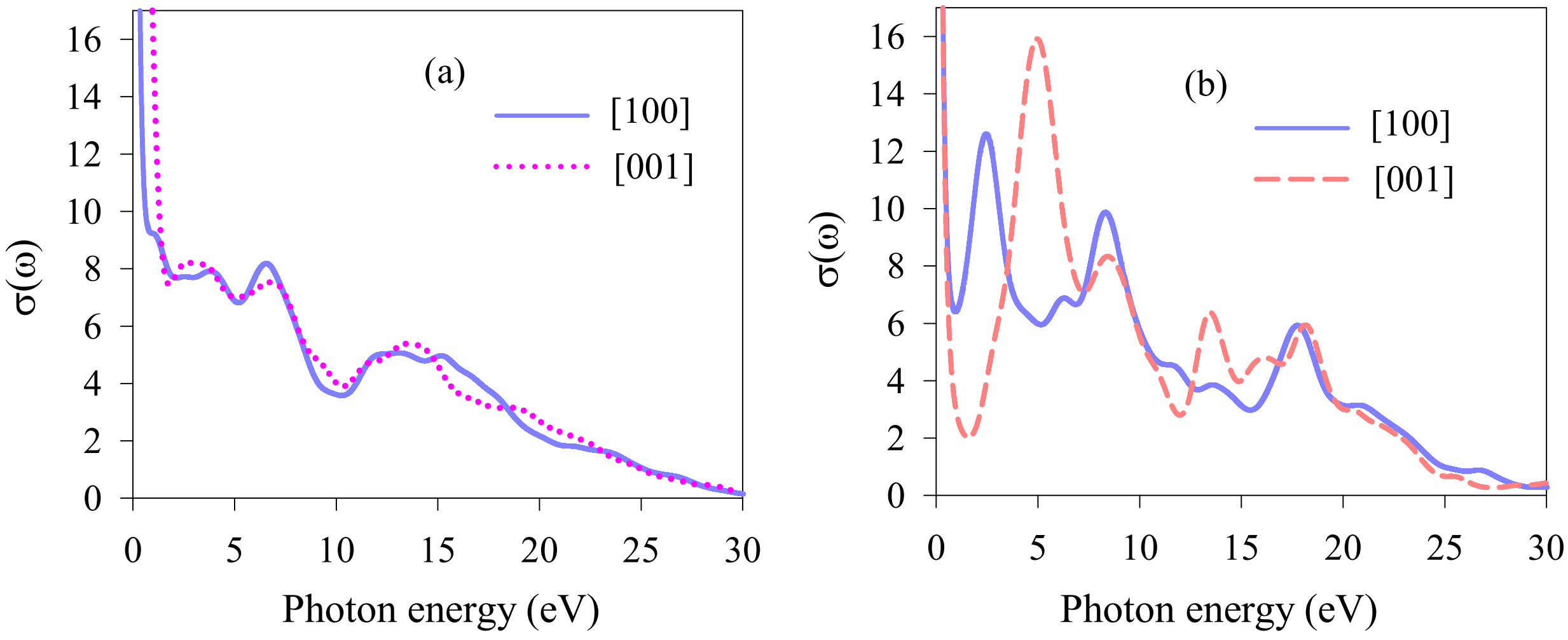

**Figure 14.** Photon energy dependence of conductivity, $\sigma(\omega)$ for (a) O-MC and (b) H-MC.

The reflectivity ($R$) is defined as the ratio of the energy of a wave reflected from a surface to the energy of the wave incident on it. Figure 15(a) and 15(b) exhibit the reflectivity spectra for both O-MC and H-MC, respectively. Unlike other optical spectra of O-MC, reflectivity shows significant optical anisotropy. Strong fall in absorption and reflectance spectra set at about 30 eV for both structures, which might be associated with the collective plasmons resonance. Reflectivity is high in the visible region for both structures; particularly for O-MC. Moreover, for energies above 10 eV in the ultraviolet region, the reflectivity shows nonselective character for O-MC. H-MC reflects ultraviolet radiation strongly for energies in the range 20 – 20 eV.

The electron energy loss spectrum ($L$) for both O-MC and H-MC are shown in Figures 16. The peaks in energy loss function spectra are associated with the plasma resonance and the corresponding frequency is called the plasma frequency ($\omega_p$) [130]. At this particular energy the charge carriers are set into collective oscillations the frequency of which depends on effective mass and concentration of the electrons. The energy loss peak appears at $\varepsilon_2 < 1$ and $\varepsilon_1 = 0$ in the high energy region [55]. The sharp energy loss peaks correspond to the trailing edges in the reflection spectra and absorption [131,132]. For instance, the peak of $L(\omega)$ is at about 28 eV corresponding the abrupt fall off in $R(\omega)$. The highest peak is found at about 27.56 eV for both polarization directions of O-MC. For H-MC, the loss peak is observed at 28.62 eV and 27.74 eV when the electric field is along [100] and [001] directions, respectively. A material becomes transparent and exhibits optical features of an insulator when the incident photon frequency is higher than the plasma frequency.

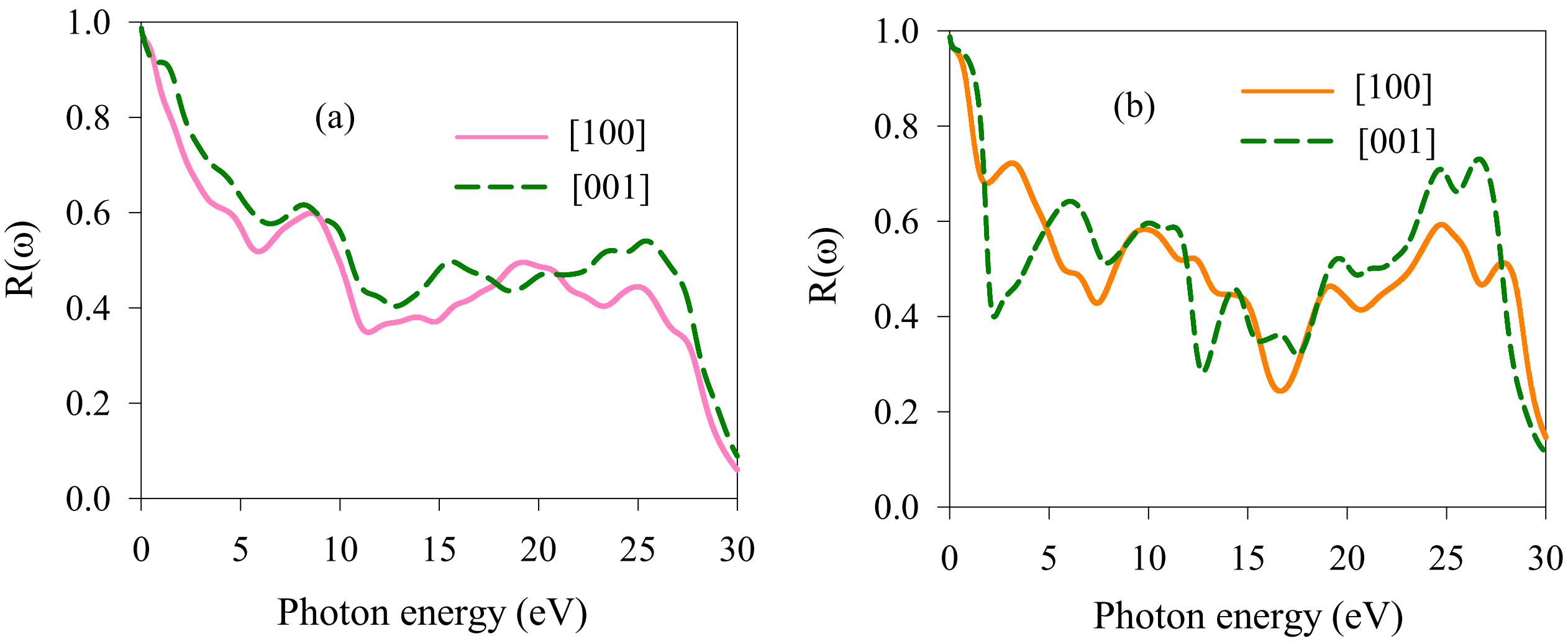

**Figure 15.** Photon energy dependence of reflection coefficient, $R(\omega)$ for (a) O-MC and (b) H-MC.

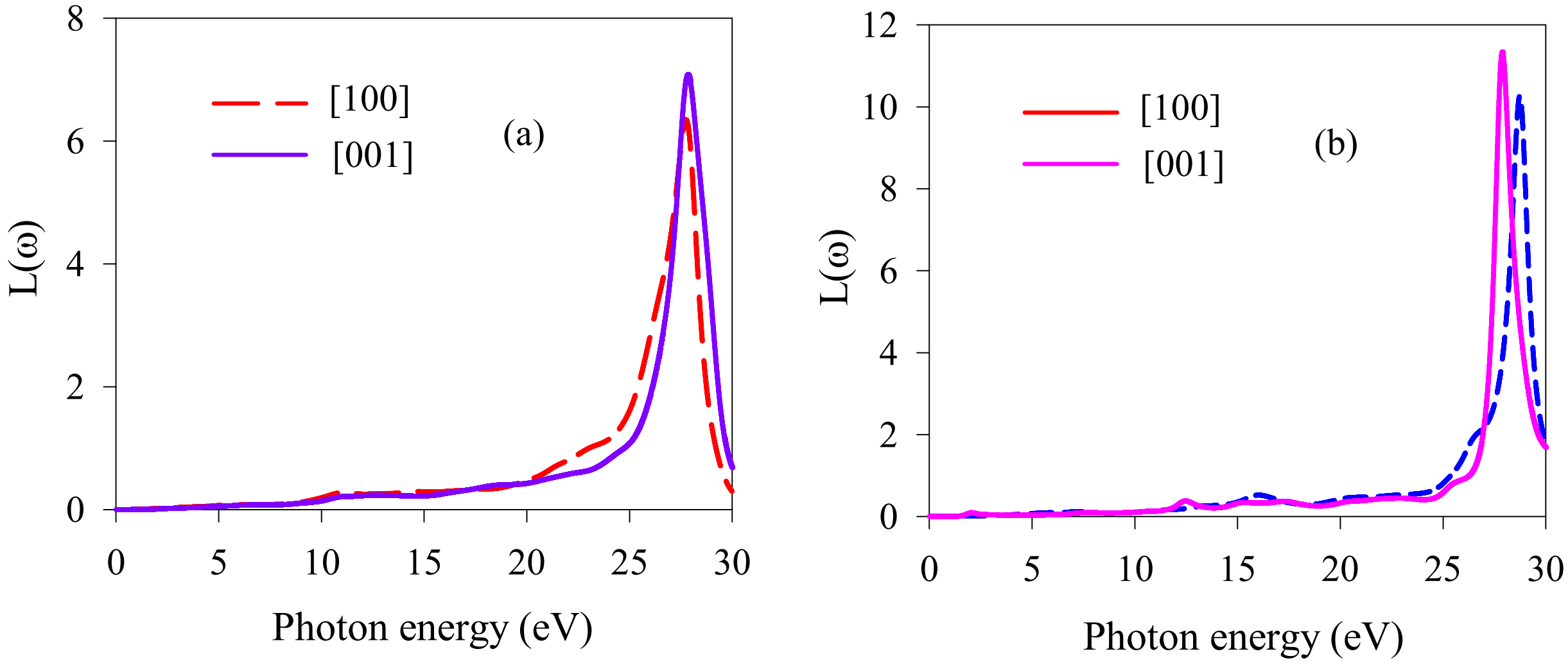

**Figure 16.** Photon energy dependence of loss function, $L(\omega)$ for (a) O-MC and (b) H-MC.

### 3.8. Bond population analysis

An often used concept in condensed matter physics is atomic charges and charge transfer in compounds. To comprehend bonding types (ionic, covalent and metallic) in O-MC and H-MC, we have used Mulliken population analysis (MPA) [59]. The charge spilling parameter and electric charges of Mo, and C species of both O-MC and H-MC are listed in Table 12. The spilling parameters for both systems are very small, indicating a good representation of the electronic bands using the LCAO basis set. A spilling parameter of 0.10 and 0.11 indicates that only approximately 0.1% of the valence charge has been missing in the projection. So, our calculations are expected to be reliable. The calculation shows that the total charge on Mo atoms is almost three times that on Ba atoms for both structures. The atomic charges of Mo and C in O-MC are 0.32*e* and -0.63*e*, respectively. Both of these values deviated from their formal charge expected from purely ionic state (Mo: +4 and C: -4). It should be remembered that the charge is transferred away from an atom if its Mulliken charge is positive and it is negative when charge is accumulated. For both structures, the calculated Mulliken charges predict that the charge is transferred from Mo atoms to C atoms. It reflects the presence of partial ionic bonding between Mo and C atoms. Covalent bondings are also present since the value of the charges deviate strongly from their canonical ionic values. The number of electrons transferred from Mo is the same as C receives. Therefore, the bonding behaviors for both structures are the combination of ionic and covalent bonds.

The effective valence charge (EVC) of an atom in a material is the difference between formal ionic charge and Mulliken charge of it (atom) [133]. It is a measure of the degree of covalency and/or iconicity. A zero value implies an ideal ionic bond while values greater than zero indicate increasing levels of covalency. The estimated effective ionic valences for both O-MC and H-MC are disclosed in Table 12. The nonzero EVC for Mo and C atoms indicates the presence of covalent bonds in both compounds.

Though Mulliken bond population analysis is a widely used method, sometimes it gives results in contradiction to chemical intuition. This is because of its sensitivity to basis set and the extended basis states never provide a natural way to quantify the local atomic properties. To get more reliable and accurate results, Hirshfeld proposed a calculation scheme called Hirshfeld population analysis (HPA) [134] which does not require a reference to basis set or their respective location. Therefore, we have also calculated Hirshfeld charge of O-MC and H-MC using the HPA. Table 12 displays comparison between Mulliken and Hirshfeld charges of both structures. The atomic charges, we got from HPA, are much lower than from MPA. But Hirshfeld charges, like Mulliken charges, also show that electrons are transferred from Mo to C atoms. The most positively charged atom is Mo, with a Hirshfeld charge of 0.34*e* in H-MC crystal. To get proper level of covalency, we have also estimated effective valence charges (EVC) for both structures from the Hirshfeld charge. All effective valences predict higher level of covalency as compared to the EVC we got from Mulliken charge.

**Table 12.** Charge spilling parameter (%), orbital charges (electron), atomic Mulliken charges (electron), formal ionic charge, EVC (electron) and Hirshfeld charge (electron) of O-MC and H-MC.

| Compounds | Charge spilling | Species | Mulliken atomic population | | | | Mulliken charge | Formal ionic charge | EVC | Hirshfeld charge | EVC |
|---|---|---|---|---|---|---|---|---|---|---|---|
| | | | *s* | *p* | *d* | total | | | | | |
| O-MC | 0.10 | Mo | 2.18 | 6.61 | 4.89 | 13.68 | 0.32 | +4 | 3.68 | 0.19 | 3.81 |
| | | C | 1.41 | 3.22 | - | 4.63 | -0.63 | -4 | 3.37 | -0.38 | 3.62 |
| H-MC | 0.11 | Mo | 2.13 | 6.49 | 4.82 | 13.44 | 0.56 | +4 | 3.44 | 0.34 | 3.66 |
| | | C | 1.38 | 3.18 | - | 4.56 | -0.56 | -4 | 3.44 | -0.34 | 3.66 |

The estimated bond overlap population, bond length, total number of each type of bond and total number of bonds in O-MC and H-MC are listed in Table 13. The calculated Mulliken bond populations define the degree of overlap of the electron clouds between two bonding atoms of solids. The overlap population of electrons between atoms is dominated by bond order. It also explains the covalent bonding strength between atoms and the bond strength per unit volume. The overlap population close to zero indicates no significant interaction between the electronic populations of the two atoms. Whereas, high positive bond population indicates a high degree of covalency in the bond. The zero overlap population represents absence of any significant interaction between the electronic populations of the two bonding atoms. In fact there are both positive and negative values of overlap populations found for O-MC. The positive (+) and negative (-) values of bond overlap populations represent the states of bonding or antibonding nature of interactions between the atoms involved, respectively [135]. Thus, the calculated values of bond overlap population indicate the presence of both bonding-type and anti-bonding-type interactions in O-MC. On the other hand, only positive overlap populations are found for H-MC indicating absence of anti-bonding state. In O-MC, there are 22 bonds which exhibit antibonding nature. A high positive bond population indicates a high degree of covalency in the bond [58]. The bond population for C-Mo bonds in H-MC is much larger than any in O-MC. Table 13 indicates the presence of both ionic and covalent bondings in our compounds.

**Table 13.** The calculated Mulliken bond overlap population of $\mu$-type bond $P^{\mu}$, bond length $d^{\mu}$ (Å), total number of $\mu$-type bond $N^{\mu}$, and total number of bond $N$ of O-MC and H-MC.

| Compounds | Bond | | $P^{\mu}$ | $d^{\mu}$ | $N^{\mu}$ | $N$ |
|---|---|---|---|---|---|---|
| O-MC | C2-Mo1<br>C1-Mo2<br>C 4-Mo7<br>C1-Mo4<br>C4-Mo5<br>C3-Mo8<br>C3-Mo6<br>C2-Mo3 | C-Mo | 0.29 | 2.083 | 8 | 66 |
| | C1-Mo6<br>C3-Mo4<br>C1-Mo8<br>C2-Mo7<br>C2-Mo5<br>C3-Mo2 | C-Mo | 0.32 | 2.092 | 8 | |

| | | | | |
|---|---|---|---|---|
| C4-Mo3<br>C4-Mo1 | | | | |
| C1-Mo3<br>C4-Mo8<br>C4-Mo6<br>C3-Mo7<br>C3-Mo5<br>C1-Mo1<br>C2-Mo4<br>C2-Mo2 | C-Mo | 0.28 | 2.110 | 8 |
| Mo4-Mo8<br>Mo2-Mo6<br>Mo1-Mo5<br>Mo3-Mo7 | Mo-Mo | 0.04 | 2.869 | 4 |
| Mo6-Mo8<br>Mo1-Mo3<br>Mo5-Mo7<br>Mo2-Mo4 | Mo-Mo | -0.19 | 2.909 | 4 |
| Mo2-Mo3<br>Mo1-Mo4<br>Mo6-Mo7<br>Mo5-Mo8 | Mo-Mo | 0.07 | 2.924 | 4 |
| C1-C3<br>C2-C4 | C-C | -0.05 | 2.976 | 2 |
| Mo4-Mo6<br>Mo3-Mo5<br>Mo2-Mo8<br>Mo1-Mo7 | Mo-Mo | 0.18 | 2.978 | 4 |
| Mo3-Mo4<br>Mo5-Mo6<br>Mo7-Mo8<br>Mo1-Mo2 | Mo-Mo | 0.19 | 3.003 | 4 |
| Mo3-Mo6<br>Mo1-Mo8<br>Mo4-Mo5<br>Mo2-Mo7 | Mo-Mo | 0.13 | 3.004 | 4 |
| C4-Mo2<br>C 3-M1<br>C1-Mo5<br>C4-Mo4<br>C3-Mo3<br>C1-Mo7<br>C2-Mo8<br>C2-Mo6 | C-Mo | -0.10 | 3.642 | 8 |
| C1-C4 | C-C | -0.00 | 3.816 | 4 |

| | | | | | | |
|---|---|---|---|---|---|---|
| | C2-C3<br>C3-C4<br>C1-C2 | | | | | |
| | Mo3-Mo8<br>Mo1-Mo6<br>Mo4-Mo7<br>Mo2-Mo5 | Mo-Mo | -0.44 | 4.202 | 4 | |
| H-MC | C2-Mo1<br>C1-Mo2<br>C2-Mo2<br>C1-Mo1 | C-Mo | 0.91 | 2.158 | 4 | 4 |

The C-Mo bonds are the strongest in both the structures and are mainly responsible for the high hardness of O-MC and H-MC compounds.

### 3.9. Thermophysical properties

Study of the thermal properties (Debye temperature, melting temperature, lattice thermal conductivity, minimum thermal conductivity, diffusion thermal conductivity, thermal expansion coefficient, etc.) of materials is important for predicting the thermodynamic stability of their structural phases and assessing their importance for variety of applications. Metals with high thermal conductivities have high electrical conductivity.

#### 3.9.1. Debye temperature

The Debye model plays a vital role in the comprehension of thermal properties of a solid phase which are related to lattice vibration. It also offers a simple but very effective method to describe the phonon contributions to the Gibbs energy of crystalline phases. Furthermore, the Debye temperature defines a line between classical and quantum-mechanical behavior of phonons and helps to separate high- and low-temperature behavior regions of a material. The thermal properties of crystal lattices such as thermal expansion, thermal conductivity, isothermal compressibility, melting point, heat capacity, and lattice enthalpy can be well described using a single parameter; the Debye temperature. At temperatures above $\theta_D$, all modes of vibrations possess the same energy, which is $k_BT$. But the higher frequency modes are considered to be frozen for temperatures below $\theta_D$ [136]. At low temperatures, the vibrational excitations in a crystal are merely dominated by the acoustic vibrations. Thus, the values of Debye temperature calculated from elastic constants and specific heat measurements coincide at low temperatures. We have calculated the Debye temperature of our compounds using the Anderson model, which is one of the standard methods [105,137]:

$$\Theta_D = \frac{h}{k_B}\left(\frac{3n}{4\pi V_0}\right)^{1/3} v_a \tag{46}$$

where, $h$ is the Planck's constant, $k_B$ is the Boltzmann's constant, $V_0$ stands for the volume of unit cell and $n$ refers to the number of atoms in a molecule. As we know, $\Theta_D$ itself is a function of temperature; it decreases with increasing temperature due to the temperature dependence of elastic constants, density and sound velocity of the solid. The calculated Debye temperatures of the compounds at 0 $K$ are listed in Table 14. Debye temperatures of orthorhombic and hexagonal structures are 369.01 $K$ and 571.86 $K$, respectively. The Debye temperature of O-MC is much lower than that of H-MC. A lower Debye temperature indicates a lower phonon thermal conductivity. Hence O-MC is expected to have lower phonon thermal conductivity and be a batter candidate material for thermal barrier coating. Debye temperature value also reflects the overall bond strength in crystals.

### 3.9.2. Lattice thermal conductivity

The lattice (phonon) thermal conductivity, $k_{ph}$, is a major parameter in a wide range of important technologies, such as in the development of new thermoelectric materials, heat sinks and thermal barrier coating[128–140]. High thermal conductivity materials are essential in microelectronic and nanoelectronic devices to achieve efficient heat removal. On the other hand, low thermal conductivity materials constitute the basis of a new generation of thermoelectric materials and thermal barrier coatings (TBC). The $k_{ph}$ of the compounds at 300 K is estimated from the formula put forward by Slack [141-143]:

$$k_{ph} = A(\gamma)\frac{M_{av}\Theta_D^3\delta}{\gamma^2 n^{2/3}T} \qquad (47)$$

In this formula, $M_{av}$ refers to the average atomic mass per atom in a compound in kg/mol, $\Theta_D$ is the Debye temperature in $K$, $\delta$ defines the cubic root of average atomic volume in m, n represents the total number of atoms in the unit cell, $T$ is the absolute temperature in $K$, $\gamma$ is the Grüneisen parameter and $A(\gamma)$, which is a function of $\gamma$, is evaluated from following equation (in W-mol/kg/m$^2$/K$^3$) [144]:

$$A(\gamma) = \frac{4{:}85628 \times 10^7}{2\left(1 - \frac{0{:}514}{\gamma} + \frac{0.228}{\gamma^2}\right)} \qquad (48)$$

The calculated room temperature lattice thermal conductivities of the compounds under study are listed in Table 14. The low lattice thermal conductivity of O-MC nicely correlates with the low Debye temperature of $\Theta_D$ = 369.01 $K$. The higher thermal conductivity indicates stronger covalent bonding. The calculated values predict that covalent bonding in H-MC is significantly stronger than that in the O-MC structure. The lattice thermal conductivity is directly related to the square root of Young's modulus of a material as: $K_L \sim \sqrt{Y}$ [145], a relation roughly holding for O-MC and H-MC.

### 3.9.3. Melting temperature

The melting temperature ($T_m$) is another essential thermophysical parameter that indicates the possibility of a material's application at higher temperatures. Crystals with higher melting temperature possess stronger atomic interaction, higher cohesive energy, higher bonding energy and lower coefficient of thermal expansion [63]. At temperatures below $T_m$ solids can be used continuously without oxidation, chemical change, and excessive distortion causing mechanical/elastic failure. It is also used to characterize organic and inorganic compounds. The melting temperature $T_m$ of O-MC and H-MC structures has been estimated from the elastic constants through the following expression [146]:

$$T_m = 354\text{ K} + \left(4.5\frac{\text{K}}{\text{GPa}}\right)\left(\frac{2C_{11} + C_{33}}{3}\right) \pm 300\text{ K} \tag{49}$$

The estimated values of $T_m$ for the studied compounds are disclosed in Table 14. The melting temperatures for O-MC and H-MC are 2589.07 $K$ and 3305.94 $K$, respectively. Like $K_{ph}$, $T_m$ of a crystal is also tightly associated with its bonding strength. H-MC has high melting temperature due to its relatively high elastic constants compared to those of O-MC. Materials used as TBC should have a high melting point so that they can withstand high operating temperatures without melting. Stiffer materials have higher melting points. A high melting point is mainly due to high heat of fusion, low entropy of fusion, or a combination of both.

### 3.9.4. Thermal expansion coefficient, dominant phonon mode and heat capacity

The thermal expansion behavior of is an intrinsic thermal property, results from anharmonic lattice vibrations. Many features of a substance, like thermal conductivity, specific heat, entropy, compressibility, temperature variation of the energy band gap in semiconductors, and electron effective mass, are related to its thermal expansion coefficient. Very low thermal expansion materials are not only important for practical uses but also have fundamental scientific interest. The thermal expansion coefficient (TEC) of a material can be calculated from shear modulus using the following expression [63,108]:

$$\alpha = \frac{1.6 \times 10^{-3}}{G} \tag{50}$$

where, $G$ is the isothermal share modulus in GPa. The TEC of a material is inversely related to its melting temperature as $\alpha \approx 0.02/T_m$ [108,146]. The TEC of crystals varies with temperature. The calculated values at 300 K are given in Table 14. Both the structures have very low thermal expansion coefficients. Low thermal expansion materials have widespread use in high anti-thermal shock applications (e.g., cookware for oven to freezer use), electronic devices, heat-engine components, spintronic devices, etc.

The wavelength at which the phonon distribution attains its peak is defined as the dominant phonon wavelength, $\lambda_{dom}$. The $\lambda_{dom}$ has significance to the thermal and electrical transport in materials. The role of long-wavelength acoustic phonons in thermal transport becomes more dominant at lower temperatures. The wavelength of the dominant phonon for a material at different temperature can be estimated from the following expression [146]:

$$\lambda_{dom} = \frac{12.566 v_a}{T} \times 10^{-12} \tag{51}$$

where, $v_a$ defines the average sound velocity in ms$^{-1}$, $T$ is the temperature in $K$. Materials with higher average sound velocity, higher shear modulus, lower density exhibits higher dominant phonon wavelength. The estimated values of $\lambda_{dom}$ at room temperature are summarized in Table 14.

The heat capacity ($C_p$) is one of the major thermodynamic quantities for material designing process. Materials with higher heat capacity exhibit higher thermal conductivity ($k$) and lower thermal diffusivity ($D$), related as $k = \rho C_P D$. The volumetric heat capacity of a substance is defined as the change in thermal energy per unit volume per Kelvin change in temperature. The volumetric heat capacity of a material can be estimated from following empirical formula [63,108]:

$$\rho C_P = \frac{3k_B}{\Omega} \tag{52}$$

here, $N = 1/\Omega$ represents the number of atoms per unit volume. The heat capacity per unit volume of O-MC and H-MC is given in Table 14.

**Table 14.** The Debye temperature $\Theta_D$ ($K$), lattice thermal conductivity $k_{ph}$ (W/m-$K$) at 300 $K$, melting temperature $T_m$ ($K$), thermal expansion coefficient α ($K^{-1}$), wavelength of dominant phonon mode $\lambda_{dom}$ (m) at 300 $K$, and heat capacity per unit volume $\rho C_P$ (J/m$^3$.$K$) of O-MC and H-MC compounds.

| Compounds | $\Theta_D$ | $k_{ph}$ | $T_m$ | $\alpha$ ($\times 10^{-6}$) | $\lambda_{dom}$ ($\times 10^{-12}$) | $\rho C_P$ ($\times 10^{6}$) | Remarks |
|---|---|---|---|---|---|---|---|
| O-MC | 369.01 | 5.71 | 2589.07 | 9.89 | 193.42 | 3.40 | This work |
| | 559.00 | - | - | - | - | - | [17]$^{Expt.}$ |
| | 580.87 | - | - | - | - | - | [33]$^{Theo.}$ |
| | 580.3 | - | - | - | - | - | [34]$^{Theo.}$ |
| H-MC | 571.86 | 44.63 | 3305.94 | 7.98 | 194.40 | 4.16 | This work |

### 3.9.5. Diffusion and minimum thermal conductivity with anisotropy

Diffusion thermal conductivity is defined as the transport of heat by diffusions. There is now a great deal of interest in the study of this phenomenon, due to its relevance to geophysics and industrial applications. Diffusions may better describe the physics of heat transfer in low thermal conductivity materials, particularly at high temperatures. The diffusion thermal conductivity, $k_{diff}$ of a material is directly related to its average energy of the lattice vibration, $\hbar\omega_{avg}$. It is also considered as an appropriate estimate of minimum thermal conductivity for materials. At high temperatures, $k_{diff} \approx k_{min}$. The diffusion thermal conductivity of a material can be estimated using following equation [147,148]:

$$k_{diff} = 0.76 n^{2/3} k_B \frac{1}{3}(2\upsilon_t + \upsilon_l) \tag{53}$$

The calculated values of the $Mo_2C$ compounds are listed in Table 15. The hexagonal structure shows better diffusion thermal conductivity than the orthorhombic structure.

The minimum thermal conductivity, $k_{min}$, carried by the atomic vibrations of any solid material (crystalline or amorphous), has technological importance. Thermal conductivity of a material decreases with increasing temperature. The theoretical minimum thermal conductivity describes the lowest limit to lattice thermal conductivity of a crystal above the Debye temperature in which the vibrational scattering rate is wavelength limited. It is also used for understanding the thermal conductivity of thin film materials. An important feature of the minimum thermal conductivity is that it is independent of the presence of defects which is largely because these defects affect phonon transport over length scales much larger than the inter-atomic spacing. The minimum thermal conductivity $k_{\mathrm{min}}$ of material described by the Clarke model is given by [146]:

$$\mathrm{k_{min}} = \mathrm{k_B} \upsilon_\mathrm{a} (\mathrm{V_{atomic}})^{-2/3} = k_B \nu_a \left(\frac{M}{n\rho N_A}\right)^{-2/3} \tag{54}$$

Here, $k_B$, $\nu_a$ and $V_{atomic}$ refer to the Boltzmann constant, average sound velocity and cell volume per atom, respectively. The estimated values of isotropic minimum thermal conductivity of our compounds are enlisted in Table 15. The Clarke model predicts that H-MC has higher minimum thermal conductivity than O-MC.

Heat propagates through a solid in three different modes: thermal vibrations of atoms, movement of free electrons in metals, and radiation. Mechanical anisotropy in materials offers anisotropy in thermal conductivity because heat transfer is dominated by elastic wave propagation. Thermal conductivity anisotropy is a subject for both fundamental and application interests. Anisotropy of thermal conductivity has important applications in particular circumstances, like in TBC and thermoelectrics. To study the anisotropy in thermal conductivity, the minimum phonon thermal conductivity along different directions can also be investigated using the Cahill's model [149]:

$$\mathrm{k_{min}} = \frac{\mathrm{k_B}}{2.48} \mathrm{n}^{2/3} (\upsilon_\mathrm{l} + \upsilon_\mathrm{t1} + \upsilon_\mathrm{t2}) \tag{55}$$

and $$n = N/V$$

where, $k_B$ is the Boltzmann constant, $n$ refers the number density of atoms and $N$ is total number of atoms in the cell having a volume $V$.

The anisotropy of minimum thermal conductivity of O-MC and H-MC along principle crystal directions is summarized in Table 15. Both compounds are almost isotropic with low $k_{min}$. Though hexagonal structure exhibit higher $k_{min}$. To compare with the Clarke model, we have also calculated isotropic minimum thermal conductivity of both compounds using Cahill's method. The estimated results are summarized in Table 15. Clearly, Clarke's model predicts smaller $k_{min}$ compared to Cahill's model. This has also been observed for other compounds [63,100,101]. Values obtained by Cahill's model should be closer to the real values than those calculated using Clarke's model. Since the Cahill's model considered the total phonon spectrum and the number density of atoms, while Clarke's model does not include the contributions to the thermal conductivity from the optical phonons [150].

**Table 15.** The number of atoms per mole of the compound $n$ ($m^{-3}$), diffusion thermal conductivity $k_{diff}$(W/m.K), minimum thermal conductivity $k_{\mathrm{min}}$ (W/m.K) of O-MC and H-MC crystals along different directions evaluated by Cahill's and Clarke's method.

| Compounds | $n$ ($10^{28}$) | $k_{diff}$ | $[100]k_{calc.}^{min}$ | $[010]k_{calc.}^{min}$ | $[001]k_{calc.}^{min}$ | $k_{min}$ | |
|---|---|---|---|---|---|---|---|
| | | | | | | Cahill | Clarke |
| O-MC | 8.21 | 1.07 | 1.75 | 1.74 | 1.72 | 1.71 | 1.23 |
| H-MC | 10.04 | 1.20 | 1.89 | - | 1.94 | 1.91 | 1.41 |

## 4. Conclusions

In the present work, we have presented and discussed a large number of hitherto unexplored elastic, electronic, thermophysical, bonding and optoelectronic properties in details of O-MC and H-MC on the basis of DFT based first-principles calculations. Both structures are elastically and dynamically stable and possess moderate level of mechanical anisotropy. Between the two crystal structures, O-MC has higher overall elastic anisotropy. Both of the structures are fairly machinable and are ductile in nature. The machinability index of $Mo_2C$ in both structures is comparable to those of many MAX phase nanolaminates [151-154]. The bondings between atoms in both compounds are mixed with notable covalent and ionic contributions. The electronic band structure and energy density of states calculations reveal metallic characteristics for both structures. The bands crossing the Fermi level of H-MC are highly dispersive compared to those of O-MC. The effective mass of the charge carriers are, therefore, significantly lower in H-MC. The estimated Coulomb pseudopotential of H-MC is also much smaller compared to that in O-MC which implies that electronic correlations are stronger in O-MC which influences the superconducting transition temperature [155,156]. The tendency of the formation of covalent bonding between the atomic species in both compounds is also obvious from their PDOS features, where significant hybridization between Mo 4*d* and C 2*p* electronic orbitals are found close to the Fermi level. The charge density distribution plots for both structures exhibit moderate direction dependency. Comparatively low values of the universal log-Euclidean index ($A^L$) predict the absence of layered characteristics and consequently the bonding strength in the structure is quite isotropic along different directions within the crystals. It is noteworthy that the orthorhombic structure shows more directional anisotropy than the hexagonal one.

The melting temperatures of both structures are high, indicating a broad range of possible applications including as TBC materials. The (lattice, diffusion, minimum) thermal conductivity of both compounds are low. All the estimated thermophysical properties of both structures are in good correspondence with each other.

O-MC and H-MC are hard materials. High hardness together with ductility and reasonable machinability make both O-MC and H-MC attractive compounds for engineering applications. The estimated values of elastic moduli, Debye temperature, minimum thermal conductivity, melting temperature imply that both compounds have significant promise to be used as a thermal barrier coating (TBC) material. The theoretically calculated Raman active modes for both the structures of $Mo_2C$ show excellent agreement with prior experimental results where available.

The optical parameters are investigated in details. The optical spectra reveal clear metallic characters consistent with band structures and DOS profiles. Both the structures possess high refractive indices in the low energy and high reflectivity in the visible range. These features are useful for optoelectronic device applications.

To summarize, we have investigated the elastic, mechanical, bonding, acoustic, thermal, bulk electronic (band, DOS, charge density distribution, electron density difference) and optoelectronic properties of O-MC and H-MC in details in this paper. The elastic, electronic, bonding, acoustic, phonon, thermal and optoelectronic properties of these crystals are studied in depth for the first time. Both compounds possess several attractive mechanical, thermal and optoelectronic features which are suitable for engineering and device applications. We are hopeful that the results obtained herein will stimulate researchers to investigate these binary carbides in greater details in near future, both theoretically and experimentally.

**Acknowledgements**
S.H.N. acknowledges the research grant (1151/5/52/RU/Science-07/19-20) from the Faculty of Science, University of Rajshahi, Bangladesh, which partly supported this work.

**Data availability**
The data sets generated and/or analyzed in this study are available from the corresponding author on reasonable request.

**Author Contributions**

M.I.N. performed the theoretical calculations, contributed to the analysis and contributed to draft manuscript. S.H.N. designed and supervised the project, analyzed the results and finalized the manuscript. Both the authors reviewed the manuscript.

**Additional Information**

**Competing Interests**

The authors declare no competing interests.